\documentclass[12pt,fleqn]{article}
\usepackage{amsmath}
\usepackage{slashed}
\usepackage{url}
\usepackage{bm}
\usepackage{graphicx}
\usepackage[utf8]{inputenc}
\usepackage[english]{babel}
\usepackage{amsmath}
\usepackage{amsfonts}
\usepackage{amssymb}
\usepackage{graphicx}
\usepackage{float}
\usepackage{hyperref}
\begin{document}
\linespread{1.5}
\renewcommand{\baselinestretch}{1.2}

\title{ Notes for Quantum Gravitational Field}
\author         {Malik AL Matwi}
\date{Department of Mathematics Science, Ritsumeikan University, 1–1 Matsugaoka 6-Chome, Otsu-city, Shiga-ken
520–2102, Japan.\\
      malik.matwi@hotmail.com}
\maketitle
\tableofcontents

\begin{abstract}
We discuss the problems of dynamics of the gravitational field and try to solve them according to quantum field theory by suggesting canonical states for the gravitational field and its conjugate field. To solve the problem of quantization of gravitational field, we assume that the quantum gravitational field $e ^I$ changes the geometry of curved spacetime $x^\mu$, and relate this changing to quantization of the gravitational field. We introduce a field $\pi _I$ and consider it as a canonical momentum conjugates to a canonical gravitational field $\tilde e^I$. We use them in deriving the path integral of the gravitational field according to quantum field theory, we get Lagrangian with dependence only on the covariant derivative of the gravitational field $e^I$, similarly to Lagrangian of scalar field in curved spacetime. Then, we discuss the case of free gravitational field. We find that this case takes place only in background spacetime approximation of low matter density; weak gravity.
Similarly, we study the Plebanski two form complex field $\Sigma ^{i}$ and derive its Lagrangian with dependence only on the covariant derivative of $\Sigma ^{i}$, which is represented in selfdual representation $\left| {\Sigma ^i } \right\rangle$.  Then, We try to combine the gravitational and Plebanski fields into one field: $K_{\mu}^i$.
Finally, we derive the static potential of exchanging gravitons between particles of scalar and spinor fields; the Newtonian gravitational potential.
\end{abstract}

Key words: Canonical gravitational field, Conjugate momentum, Path integral, Free gravitational propagator, Lagrangian of Plebanski field, Combination of Plebanski and gravitational fields.

\section*{Introduction}

We search for conditions allow us to consider the gravitational field as dynamical field. The problem of dynamics in general relativity is that spacetime is itself dynamical, that it acts on matter. But to define dynamical gravitational field, we have to fix coordinates $ x^\mu$ on spacetime manifold $M$, and regard the gravitational field $ e^I$ as dynamical field which interacts with matter while keeping the charts $ x^\mu$ fixed. We will see that in background spacetime the gravitational field can be considered as a free field similarly to free quantum fields in flat spacetime.\\

To study dynamics of gravitational field which changes geometry of spacetime, we write the spin connection $\omega^{IJ}$ of local Lorentz frame as $ \Omega^{IJ} + B^{IJ}$ by using a reference connection $B^{IJ}$ and a tensor $ \Omega^{IJ}$. We relate this changing in the geometry to fluctuation of the gravitational field. It is like to say that the quantum gravitational field consists of quanta (like $\delta e ^I$), each of them contributes to spin connection changes $\delta\omega^{IJ}$, so dynamical changes in geometry of spacetime. Those changes are seen in the tensor $ \Omega^{IJ}$. \\

To define dynamical variables and so canonical states, we introduce an anti-symmetric tensor $\pi^{IJK}$ by the formula ${\Omega }^{IJ}= \pi {_K} ^{IJ} e^K $ and a field $\pi _I$ to satisfy $\pi ^{IJK}  = \pi _L \varepsilon ^{LIJK}$. We consider $\pi _I$ as a canonical momentum conjugates to a canonical gravitational field $\tilde e^I =  ee_{\mu}^I n^{\mu}$, where $n^{\mu}$ is normal to 3D closed surface $\delta M$, which is embedded in arbitrary curved 4D spacetime manifold $M$. The closed surface $\delta M$ is parameterized by three parameters $X^1, X^2$ and $ X^3$. We consider them in a certain gauge as spatial part of the local-Lorentz frame $X^I: I=0, 1, 2, 3$. We will see that the path integral of the gravitational field is independent of this gauge.\\
Therefore, the exterior derivative operator on the surface $\delta M(X^1, X^2, X^3)$ leads to a changing along the normal of that surface in direction of the time $dX^0$ . That allows the 3D surface $\delta M(X^1, X^2, X^3)$ to extend, and so obtaining the local-Lorentz frame $(X^0, X^1, X^2, X^3)$. With that the gravitational field propagates from one surface to another by the extension of those surfaces.\\

So we can introduce canonical states $\left| {\tilde e^I } \right\rangle$ and $\left| {\pi ^I } \right\rangle$ represented in local Lorentz frame and satisfy $\left\langle {{\tilde e^I }} \mathrel{\left | {\vphantom {{\tilde e^I } {\pi _I }}} \right. \kern-\nulldelimiterspace} {{\pi _I }} \right\rangle _{\delta M}  = \exp i\int_{\delta M} {\tilde e^I (X)\pi _I (X)d^3 X}$. We use them in deriving the path integral of the gravitational field. We find that there is no propagation on the dynamical spacetime $x^\mu$. But in the background spacetime, the gravitational field propagates freely like the electromagnetic ans scalar fields.\\

We start with using our formula $\omega^{IJ}=\Omega^{IJ} + B^{IJ}$ in Einstein-Hilbert Lagrangian $L(e,\omega ) = c {\varepsilon _{IJKL} e^I  \wedge e^J  \wedge R^{KL} (\omega )}$, we get 
\[
L(e,\Omega, B ) =c\varepsilon _{IJKL} e^I  \wedge e^J  \wedge \left(\Omega\wedge \Omega +d\Omega +dB+\Omega\wedge B+B\wedge\Omega+B\wedge B\right)^{KL}.
\]
Because $\Omega^{IJ}$ is tensor, the part $c\varepsilon _{IJKL} e^I  \wedge e^J  \wedge \left(\Omega\wedge \Omega\right)^{KL}$ of this Lagrangian is scalar. If we use this part in path integral for the states $\left| {\tilde e^I } \right\rangle$ and $\left| {\pi ^I } \right\rangle$, with eq.(1.4), we get the Lagrangian 
\[
L(e,\Omega, B) =c\varepsilon _{IJKL} e^I  \wedge e^J  \wedge \left(\Omega\wedge \Omega \right)^{KL}+\pi _I D\tilde e^I d^3 X.
\]
Similarly to the path integral of one particle in one dimension for the states $\left| {x, t } \right\rangle=e^{iHt}\left| {x } \right\rangle$, with using the formula 
  $ \left\langle {x}\mathrel{\left | {\vphantom {x p}} \right. \kern-\nulldelimiterspace} {p} \right\rangle  = \left( {2\pi } \right)^{ - 1/2} \exp (ipx)$, which is
\[
\left\langle {{x_2 ,t_2 }} \mathrel{\left | {\vphantom {{x_2 ,t_2 } {x_1 ,t_1 }}} \right. \kern-\nulldelimiterspace} {{x_1 ,t_1 }} \right\rangle  = \int {DxDp\exp i\int\limits_{t_1 }^{t_2 } {dt\left( {p\dot x - H} \right)} } .
\]
The role of the term $c\varepsilon _{IJKL} e^I  \wedge e^J  \wedge \left(\Omega\wedge \Omega\right)^{KL}$ in our Lagrangian is same role of the Hamiltonian $H$ in this path integral, while the term $\pi _I D\tilde e^I d^3 X$ is similar to the term $p\dot x$. Therefore, we suggest the equality 
\[
c\varepsilon _{IJKL} e^I  \wedge e^J  \wedge \left(d\Omega +dB+\Omega\wedge B+B\wedge\Omega+B\wedge B\right)^{KL}=\pi _I D\tilde e^I d^3 X.
\]
We try to show that there is at least one solution for this equation at end of section 1, (\ref{eq:b12}). The covariant derivative $D=d+\omega$ includes $\omega$, so this equality does not decrease the number of variables, $e^I$ and $ \omega^{IJ}$. \\

Because the exterior derivative $D\tilde e^I$ allows expanding of the 3D closed surface $\delta M$, we consider the term $\pi _I D\tilde e^I d^3 X$ as a kinetic energy which relates to the expansion of those surfaces, while consider the term $c\varepsilon _{IJKL} e^I  \wedge e^J  \wedge \left(\Omega\wedge \Omega\right)^{KL}$ as gravitational energy of $\tilde e^I$ on the surface $\delta M$ on which the states $\left| {\tilde e^I } \right\rangle$ and $\left| {\pi ^I } \right\rangle$ are defined, eq.(1.4). Finally, we get the Lagrangian:
\[
L (e,\omega)= \frac{1}{{4\pi G}}{\frac{1}{4}}\left( { - D_\mu  e_I^\nu  D^\mu  e_\nu ^I  + D_\mu  e_I^\nu  D_\nu  e^{I\mu } } \right)ed^4 x,
\]
where the covariant derivative $D$ is defined in $DV^I=dV^I+\omega^I{_J}\wedge V^J$, here we use the usual spin connection $\omega^{IJ}$ not our $\Omega^{IJ} + B^{IJ}$. Actually, we use $\Omega^{IJ} + B^{IJ}$ just for getting this Lagrangian. Therefore, we consider the states $\left| { e^{I}_\mu } \right\rangle$ and $\left| {\omega^{IJ}_\mu } \right\rangle$ for the path integral of this Lagrangian.
\\

We will show that this Lagrangian gives Einstein's equation in vacuum \ref{eq:b11}, and write this Lagrangian using Riemann curvature tensor ${(R_{\mu\nu})}_{IJ}$. For ${D_\mu}e^{I\mu} =0$, we get
\[
Ld^4 x \to \frac{1}{{48c}}\frac{1}{2}\left( {   e_I^\mu  D^2 e_\mu ^I  +  e^{I\mu} e^{J \nu}{(R_{\mu\nu})}_{IJ}   } \right)ed^4 x.
\]

We use same method for Plebanski two form real field $\Sigma^{IJ}_{\mu\nu}$ and get the Lagrangian
\[
L(e, \Sigma )=-4c'\left( {D_\mu  \Sigma _{IJ}^{\nu \rho } } \right)\left( {D^\mu  \Sigma _{\nu \rho }^{IJ} } \right)ed^4 x.
\]
Then, we write this Lagrangian using the complex Plebanski selfdual two-form $\Sigma^{i}_{\mu\nu}$, and search for the reality conditions.\\

\section{ Lagrangian and path integral of quantum gravitational field}

We use the indices notation $\mu, \nu, ...=0, 1, 2, 3$ for 4D spacetime tangent space on arbitrary spacetime manifold $M$, and the indices notation $I, J, ...=0, 1, 2, 3$ for 4D Lorentz tangent space with the metric $[- + + +]$. At each point $x^\mu$ in the manifold $M$, we define gravitational field $e^I=e^I_\mu (x) dx^\mu$, and spin connection $ \omega^{IJ}(x)=\omega^{IJ}_\mu (x) dx^\mu$ with values in Lie algebra of Lorentz group $SO(3, 1)$. The spin connection defines covariant derivative $D_\mu$ on all fields that have Lorentz indices $(I, J, ...)$[1, 2]:
\[
D_\mu v^I=\partial_\mu v^I + \omega^{I}_{\mu J} v^J.
\]
The gravitational field determines compatible spin connection by
\[
D e^I=d e^I + \omega{^I}_{ J}\wedge e^J=0.
\]
Actually we consider this solution for the spin connection as a classical limit. So, the quantum fluctuation of gravitational field $e^I$ violates this formula. \\

The spin connection transforms under local Lorentz transformation $ L(x){^{I}}_J$, in a matrix notation, as[3]
\[
\omega ' =L \omega L ^{ - 1}  +L d L ^{ - 1} \text{ }\text{ }or \text{ }\text{ } \omega '_\mu  dx^\mu   = L \omega _\mu   L ^{ - 1} dx^\mu  + L \partial _\mu  L ^{ - 1} dx^\mu .
\]
 We write the connection $\omega$ as 
\begin{equation}\numberwithin{equation}{section}
 \omega _\mu  dx^\mu   = \Omega _\mu  dx^\mu + B_\mu dx^\mu,
\end{equation}
where $B^{IJ}$ is a reference connection $\left(\text{consider it as  spin connection in the vacuum}\right)$, and $\Omega^{IJ}_\mu  dx^\mu$ is a tensor, so it transforms covariantly under local Lorentz transformation: $L\Omega L^{-1}=\Omega'$. Therefore,
\[
 L\left(\omega  - B\right)L ^{ - 1}=\omega'  - B'
\]
 which yields 
\begin{equation}\numberwithin{equation}{section}
 B'=LBL^{-1}+LdL^{ - 1} \text{ } with \text{ }\omega'=L\omega L^{-1} + LdL^{ - 1},
\end{equation}
for $B=\left( B^{IJ} \right)$ and $\omega=\left( \omega^{IJ} \right)$,\\
and also yields
\begin{equation}\numberwithin{equation}{section}
 B'=LBL^{-1}-(dL)L^{ - 1} \text{ } with \text{ }\omega'=L\omega L^{-1} - (dL)L^{ - 1} ,
\end{equation}
for $B=\left( B{^I}_J \right)$ and $\omega=\left( \omega{^I}_J \right)$.\\
\\
 For dynamical gravitational field, let us suggest an anti-symmetry tensor $\pi ^{IJK}$ to satisfy
\[
{\Omega }^{IJ}= \pi {_K} ^{IJ} e^K .
\]
We consider it as a conjugate momentum represented in the local Lorentz frame and acts on its vectors. Therefore, we consider it as dynamical operator. Let us introduce a field $\pi _I $ to satisfy
\[
\pi ^{IJK}  = \pi _L \varepsilon ^{LIJK}.
\]
The element
\[
e\varepsilon _{\mu \nu \rho \sigma } dx^\nu   \wedge dx^\rho   \wedge dx^\sigma  /3! = d^3 x_\mu
\]
is a co-vector, as $\partial _\mu$, therefore 
\[
\pi _K e^{K\mu } d^3 x_\mu   = \pi _K e^{K\mu } e\varepsilon _{\mu \nu \rho \sigma } dx^\nu   \wedge dx^\rho   \wedge dx^\sigma  /3!
\]
is invariant under local Lorentz transformation $V^I  \to L_J^I (x)V^J$ and under arbitrary changing of the coordinates $dx^\mu   \to \Lambda {^\mu} {_\nu}   (x)dx^\nu$.
\\
Let us integral it over 3D closed surface $\delta M$ in arbitrary curved spacetime manifold $M$, and introduce the phase
\[
\exp i\oint_{\delta M} {\pi _I e^{I\mu } e\varepsilon _{\mu \nu \rho \sigma } dx^\nu   \wedge dx^\rho   \wedge dx^\sigma  /3!}
\]
in which we can define a canonical gravitational field $\tilde e^I$ via
\[
\tilde e^I d^3 X = \tilde e^I dX^1\wedge dX^2 \wedge dX^3  \equiv e^{I\mu } e\varepsilon _{\mu \nu \rho \sigma } dx^\nu   \wedge dx^\rho   \wedge dx^\sigma  /3! ,
\]
or
\[
\tilde e^I =  ee^{I\mu} n_{\mu}{\left(X^i\right)},
\]
where 
\[
n_{\mu}{(X^i)}= \frac{{\varepsilon _{\mu \nu \rho \sigma } }}{{3!}}\frac{{\partial x^\nu  }}{{\partial X^i }}\frac{{\partial  x^\rho  }}{{\partial X^j }}\frac{{\partial  x^\sigma  }}{{\partial X^k }}\frac{{\varepsilon ^{ijk} }}{{3!}}
\]
 is the normal to the surface $\delta M(X^1, X^2, X^3)$.
So, we get
\[
 \exp i\oint_{\delta M} {\pi _I \tilde e^I d^3 X},
\]
with the parameters $X^I: I=i=1,2,3$ parameterize the closed 3D surface $\delta M$. As mentioned before, in a certain gauge, we consider those parameters as spatial part of the local-Lorentz frame $X^I: I=0,1,2,3$. Therefore, the exterior derivative on the surface $\delta M$ is along the direction of the time $dX^0$, which is the direction of the normal to the surface $\delta M(X^1, X^2, X^3)$. We will see that the result of the path integral is independent of this gauge.
\\
By comparing the previous formula with
\[
\left\langle {\phi } \mathrel{\left | {\vphantom {\phi  \pi }} \right. \kern-\nulldelimiterspace} {\pi } \right\rangle  = \exp i\int {d^3 X\phi (X)\pi (X)} /\hbar,
\]
which is a canonical formula in the scalar field theory on flat spacetime[4], for $\hbar=1$, we suggest canonical states $\left| {\tilde e^I } \right\rangle$ and $\left| {\pi ^I } \right\rangle$ with
\begin{equation}\numberwithin{equation}{section}\label{eq:1}
\left\langle {{\tilde e^I }} \mathrel{\left | {\vphantom {{\tilde e^I } {\pi _I }}} \right. \kern-\nulldelimiterspace} {{\pi _I }} \right\rangle _{\delta M}  = \exp i\int_{\delta M} {\tilde e^I (X)\pi _I (X)d^3 X} ,
\end{equation}
where $\pi _I$ is canonical momentum conjugates to $\tilde e^I$.
Let us write this formula on the surface $\delta M(X^i) $ as
\[
\left\langle {{\tilde e^I }}
 \mathrel{\left | {\vphantom {{\tilde e^I } {\pi _I }}}
 \right. \kern-\nulldelimiterspace}
 {{\pi _I }} \right\rangle _{\delta M}  = \mathop \prod \limits_{n,I} \left\langle {{\tilde e^I \left( {x_n  + dx_n } \right)}}
 \mathrel{\left | {\vphantom {{\tilde e^I \left( {x_n  + dx_n } \right)} {\pi _I \left( {x_n } \right)}}}
 \right. \kern-\nulldelimiterspace}
 {{\pi _I \left( {x_n } \right)}} \right\rangle _{\delta M},
\]
with
\[
\left\langle {{\tilde e^I \left( {x_n  + dx_n } \right)}}
 \mathrel{\left | {\vphantom {{\tilde e^I \left( {x_n  + dx_n } \right)} {\pi _I \left( {x_n } \right)}}}
 \right. \kern-\nulldelimiterspace}
 {{\pi _I \left( {x_n } \right)}} \right\rangle _{\delta M}  = \exp iU_J^I(x_n  + dx_n,x_n )\tilde e^J (x_n  + dx_n )\pi _I (x_n )d^3 X  ,
\]
where 
\[
U(x_n  + dx_n,x_n )=I+\omega_\mu(x_n)dx^\mu+O(dx^2)\in SO(3,1)
\] 
is parallel transport that produces covariant exterior derivative as 
\[
U_J^I(x_n  + dx_n,x_n )\tilde e^J (x_n  + dx_n )-\tilde e^I (x_n   )=D\tilde e^I (x_n   )+O(dx^3),
\]
with
\[
D\tilde e^I  = d\tilde e^I + \omega {^I} {_J} \wedge \tilde e^J.
\]
In general, for two points in adjacent surfaces $\delta M_1$ and $\delta M_2$, let us rewrite it as
\begin{equation}\numberwithin{equation}{section}\label{eq:2}
\left\langle {{\tilde e^I \left( {x_n  + dx_n } \right)}}
 \mathrel{\left | {\vphantom {{\tilde e^I \left( {x_n  + dx_n } \right)} {\pi _I \left( {x_n } \right)}}}
 \right. \kern-\nulldelimiterspace}
 {{\pi _I \left( {x_n } \right)}} \right\rangle  = \exp iU_J^I(x_n  + dx_n,x_n )\tilde e^J (x_n  + dx_n )\pi _I (x_n )d^3 X .
\end{equation}
The exterior covaraint derivative 
\[
U_J^I(x_n  + dx_n,x_n )\tilde e^J (x_n  + dx_n )-\tilde e^I (x_n   )=D\tilde e^I (x_n   )+O(dx^3)
\]
is along the time $dX^0$ in the direction of the normal $n^{\mu}{(X^i)}$ to the surface $\delta M_1$. It allows the extension of this surface: $\delta M(X^1, X^2, X^3) \to M(X^0, X^1, X^2, X^3)$. This leads to the propagation of those surfaces.\\

We need to make $\hat ed^4 \hat x$ commute with $\hat {\tilde e}{^I} d^3 X$. For this purpose, we write
\begin{align*}
 - ed^4  x &=  ed x^\mu   \wedge \varepsilon _{\mu \nu \rho \sigma } d x^\nu   \wedge d x^\rho   \wedge d x^\sigma  /4! \\
 {\rm{        }} &=  ed x^\mu   \wedge \frac{{\varepsilon _{\mu \nu \rho \sigma } }}{{4!}}\frac{{\partial x^\nu  }}{{\partial X^i }}\frac{{\partial  x^\rho  }}{{\partial X^j }}\frac{{\partial  x^\sigma  }}{{\partial X^k }}\frac{{\varepsilon ^{ijk} }}{{3!}}d^3 X = \frac{1}{4} ed x^\mu   n_\mu  d^3 X .
\end{align*}
The indices $\emph{i, j}$ and ${  k}$ in our gauge are the local-Lorentz frame indices for $I=1,2,3$. We can rewrite it as
\[
-ed^4 x = \frac{1}{4}edx^\mu  n_\mu  d^3 X = \frac{1}{4}e\frac{{\partial x^\mu  }}{{\partial X^0 }}n_\mu  d^3 XdX^0  = \frac{1}{4}ee_0^\mu  n_\mu  d^3 XdX^0 .
\]
Comparing it with the term
\[
\tilde e^I d^3 X = e^{I\mu } e\varepsilon _{\mu \nu \rho \sigma } dx^\nu   \wedge dx^\rho   \wedge dx^\sigma  /3! = ee^{I\mu } n_\mu  d^3 X,
\]
we find that it commutes with it:
\[
\left[ {\hat e\hat e^{I\mu } \hat n_\mu  d^3 X,\hat e\hat e_0^\mu  \hat n_\mu  d^3 XdX^0 } \right] = 0 \to \left[ {\hat {\tilde e}{^I} d^3 X,\hat ed^4 \hat x} \right] = 0,
\]
where $\left[ {\hat e_\mu ^I ,\hat e_\nu ^J } \right] = 0$ .
Thus, the operator $\hat ed^4 \hat x$ takes eigenvalues when it acts on the states $\left| {\tilde e^I } \right\rangle$.\\

The Einstein's action for gravity written in the first order formalism is expressed by[5]
\[
S(e,\omega ) = \frac{1}{{16\pi G}}\int {\varepsilon _{IJKL} \left( {e^I  \wedge e^J  \wedge R^{KL} (\omega ) + \lambda e^I  \wedge e^J  \wedge e^K  \wedge e^L } \right)}.
\]
Let us consider only the first term:
\begin{equation}\numberwithin{equation}{section}\label{eq:c3}
S(e,\omega ) = c\int {\varepsilon _{IJKL} e^I  \wedge e^J  \wedge R^{KL} (\omega )},
\end{equation}
where $C$ is constant. The Riemann curvature here is
\[
R^{KL} (\omega ) = d\omega ^{KL}  + \omega {^K }{_M} \wedge \omega ^{ML}.
\]
Inserting the formula we suggested before:
\begin{equation}\numberwithin{equation}{section}\label{eq:c4}
\omega ^{IJ} =\Omega ^{IJ} +B^{IJ} \text{ },\text{ }\Omega ^{IJ}= \pi {_K} ^{IJ} e ^{K},
\end{equation}
the action becomes
\begin{equation}\numberwithin{equation}{section}
 S(e,\pi )= c\int \varepsilon _{IJKL} e^I  \wedge e^J  \wedge \left( d\Omega +dB+\Omega\wedge \Omega+\Omega\wedge B+B\wedge\Omega+B\wedge B\right)^{KL}.
\end{equation}
As we suggested before that $\Omega ^{IJ}$ transforms covariantly, so the remaining term $d\Omega +dB+\Omega\wedge B+B\wedge\Omega+B\wedge B$ also transforms covariantly, this is because the Lagrangian transforms covariantly.
\\
Let us test it by using the formulas eq.(1.1), eq.(1.2) and eq.(1.3), it becomes
\[
\begin{array}{l}
 d\left(L\Omega L^{-1}\right) + d\left(LBL^{-1}+LdL^{ - 1}\right) +L\Omega L^{ - 1}\wedge\left(LBL^{-1}+LdL^{ - 1}\right) \\ 
\\
 +\left(LBL^{-1}-(dL)L^{ - 1}\right)\wedge L\Omega L^{ - 1} +\left(LBL^{-1}-(dL)L^{ - 1}\right)\wedge\left(LBL^{-1}+LdL^{ - 1}\right).\\ 
 \end{array}
\]
Expanding it:
\[
\begin{array}{l}
 \left(dL\right)\wedge\Omega L^{-1} +L\left(d\Omega\right) L^{-1}-L\Omega\wedge\left(dL^{-1}\right) + \left(dL\right)\wedge BL^{-1}+L\left(dB\right)L^{-1}-LB\wedge dL^{-1}\\
\\
+\left(dL\right)\wedge dL^{ - 1}+L\Omega L^{ - 1}\wedge LBL^{-1}+L\Omega L^{ - 1}\wedge LdL^{ - 1} +LBL^{-1}\wedge L\Omega L^{ - 1} \\
\\
-(dL)L^{ - 1}\wedge L\Omega L^{ - 1}+LBL^{-1}\wedge LBL^{-1} -(dL)L^{ - 1} \wedge LBL^{-1} + LBL^{-1} \wedge LdL^{ - 1} \\
\\
- (dL)L^{ - 1} \wedge LdL^{ - 1},\\
 \end{array}
\]
so
\[
\begin{array}{l}
 \left(dL\right)\wedge\Omega L^{-1} +L\left(d\Omega\right) L^{-1}-L\Omega\wedge\left(dL^{-1}\right) + \left(dL\right)\wedge BL^{-1}+L\left(dB\right)L^{-1}-LB\wedge dL^{-1}\\
\\
+\left(dL\right)\wedge dL^{ - 1}+L\Omega \wedge BL^{-1}+L\Omega \wedge dL^{ - 1} +LB\wedge \Omega L^{ - 1} -(dL)\wedge \Omega L^{ - 1}\\
\\
+LB\wedge BL^{-1} -(dL) \wedge BL^{-1} +LB \wedge dL^{ - 1} - (dL)\wedge dL^{ - 1},\\
 \end{array}
\]
it becomes
\[
L\left(d\Omega\right) L^{-1} +L\left(dB\right)L^{-1}+L\Omega \wedge BL^{-1} +LB\wedge \Omega L^{ - 1} +LB\wedge BL^{-1} .\\
\]
As expected, it transforms covariantly. Therefore, we can choose the equality
\begin{equation}\numberwithin{equation}{section}\label{eq:3}
 c\varepsilon _{IJKL} e^I  \wedge e^J  \wedge\left(d\Omega +dB+\Omega\wedge B+B\wedge\Omega+B\wedge B\right)^{KL}=\pi _I D\tilde e^I d^3 X,
\end{equation}
where $D\tilde e^I$ is the exterior covariant derivative of $\tilde e^I$ along the normal of the surface $\delta M(X^1, X^2, X^3)$ as mentioned before. We try to show that there is at least one solution for this equation at end of this section (\ref{eq:b12}), with $\pi^I$ given in \ref{eq:6}.\\

We postulated this equality because if we use the action $c\varepsilon _{IJKL} e^I  \wedge e^J  \wedge \left(\Omega\wedge \Omega\right)^{KL}$ in the path integral for the states $\left| {\tilde e^I } \right\rangle$ and $\left| {\pi ^I } \right\rangle$, with \ref{eq:1} and \ref{eq:2}, we get the action
\begin{equation}\numberwithin{equation}{section}\label{eq:4}
 S(e,\pi )= \int \left(c\varepsilon _{IJKL} e^I  \wedge e^J  \wedge \left(\Omega\wedge \Omega \right)^{KL}+\pi _I D\tilde e^I d^3 X\right).
\end{equation}
For $\pi^I =0$, with the formulas $\omega=\Omega+B$ and $\Omega ^{IJ}= \pi {_K} ^{IJ} e ^{K}$, thus \ref{eq:3} becomes
\[
  \varepsilon _{IJKL} e^{I}  \wedge e^{J}  \wedge\left(dB+B\wedge B\right)^{KL}=0.
\]
This is the Lagrangian of general relativity in the vacuum (for $L=H=0$), therefore we can consider the reference connection $B$ as spin connection in the vacuum.\\

Now, we get the Lagrangian \ref{eq:4} by using the path integral for the action $S_1$:
\begin{equation}\numberwithin{equation}{section}\label{eq:1a}
S_1= c\int \varepsilon _{IJKL} e^I  \wedge e^J  \wedge \left(\Omega\wedge \Omega \right)^{KL}= c\int \varepsilon _{IJKL} e^I  \wedge e^J  \wedge \Omega{^K}_M \wedge \Omega ^{ML}.
\end{equation}
We use our assumption $\Omega ^{IJ}= \pi {_K} ^{IJ} e ^{K}$ in this action, we get
\[
S_1= c\int \varepsilon _{IJKL}  e^I  \wedge e^J  \wedge\left( {\pi{ _{K_1 }}{^K}{ _M }} \right) e^{K_1 }  \wedge\left( {\pi{ _{K_2} } ^{ML} } \right)e^{K_2 }.
\]
 Making the replacement
\[
e^I  \wedge e^J  \wedge e^{K_1 }  \wedge e^{K_2 }  \to \varepsilon ^{IJK_1 K_2 } e^0  \wedge e^1  \wedge e^2  \wedge e^3,
\]
we get
\[
S_1= c\int\varepsilon _{IJKL} \left( {\pi {_{K_1 }} {^K}{_M} } \right)\left( {\pi{ _{K_2} } ^{ML} } \right)\varepsilon ^{IJK_1 K_2 } {e^0  \wedge e^1  \wedge e^2  \wedge e^3 }.
\]
Inserting the relation $\pi ^{IJL}  = \pi _K \varepsilon ^{KIJL}$  we imposed before, and using $\varepsilon _{IJKL} \varepsilon ^{IJK_1 K_2 }  =  - 2\left( {\delta _{K}^{K_1} \delta _{L}^{K_2 }  - \delta _{L}^{K_1} \delta _{K}^{K_2 } } \right)$, we obtain
\[
S_1 =c\int 2\pi ^I \varepsilon _{ILMK} \pi _J \varepsilon ^{JLMK}  e^0  \wedge e^1  \wedge e^2  \wedge e^3,
\]
and using $\varepsilon _{ILMK} \varepsilon ^{JLMK}  =  - 6\delta _{I}^J$, it becomes
\[
S_1 =- 12c\int  {\pi _I}{ \pi ^I }e^0  \wedge e^1  \wedge e^2  \wedge e^3,
\]
so
\begin{equation}\numberwithin{equation}{section}\label{eq:2a}
S_1 = -12c\int {\pi ^2 ed^4 x}.
\end{equation}
This action is scalar, so we can use it in the path integral for the states $\left| {\tilde e^I } \right\rangle$. We consider $ -12c\int {\pi ^2 ed^4 x}$ as self-energy of $\tilde e^I$ on the closed surface $\delta M$ on which the states $\left| {\tilde e^I } \right\rangle$ and $\left| {\pi^I } \right\rangle$ are defined with \ref{eq:1} and \ref{eq:2}. As we saw before, in our gauge, the operator $\hat ed^4 x$ takes eigenvalues when it acts on the states $\left| {\tilde e^I } \right\rangle$. Using \ref{eq:2}, we get the amplitude
\begin{align*}
& \left\langle {\tilde e^I \left( {x + dx} \right)} \right|e^{i\hat S_1} \left| {\pi _I \left( x \right)} \right\rangle = \left\langle {\tilde e^I \left( {x + dx} \right)} \right|e^{-i12c\hat \pi ^2 \hat ed^4  x} \left| {\pi _I \left( x \right)} \right\rangle  \\
\\
 &= \exp \left( {-i12c\pi ^2\left( {x} \right) e\left( {x + dx} \right)d^4 x + iU_J^I(x  + dx,x)\tilde e^J (x+ dx)\pi _I (x)d^3 X} \right) \\ \text{ }\text{ }\text{ }
\\
  &\to \exp \left( {-i12c\pi ^2\left( {x} \right) e\left( x \right)d^4 x + iU_J^I(x+ dx,x )\tilde e^J (x+ dx )\pi _I (x)d^3 X} \right).\\
\end{align*}
The amplitude of the propagation between two points $x$ and $x+dx$ of adjacent surfaces $\delta M_1$ and $ \delta M_2$ is
\[
\begin{array}{l}
\left\langle {\tilde e_I \left( {x + dx} \right)} \right|e^{-ic12\hat \pi ^2 \hat ed^4  x} \left| {\tilde e^I \left( x \right)} \right\rangle _{\delta M_1  \to \delta M_2 }  \\
\\
= \int {\mathop \prod \limits_I d\pi ^I \left\langle {\tilde e_I \left( {x + dx} \right)} \right|e^{-ic12\hat \pi ^2 \hat ed^4  x} \left| {\pi ^I \left( x \right)} \right\rangle _{\delta M_1  \to \delta M_2 } \left\langle {{\pi _I \left( x \right)}}
 \mathrel{\left | {\vphantom {{\pi _I \left( x \right)} {\tilde e^I \left( x \right)}}}
 \right. \kern-\nulldelimiterspace}
 {{\tilde e^I \left( x \right)}} \right\rangle _{\delta M_1 } }\\
\\

 {\rm{ = }}\int {\mathop \prod \limits_I d\pi ^I \exp \left[ {-i12c\pi ^2 \left( x \right)e\left( {x + dx} \right)d^4 x +i U^I_J\tilde e^J (x + dx)\pi _I (x)d^3 X} \right]\exp \left( { - i\tilde e^I (x)\pi _I (x)d^3 X} \right)}  \\
\\
 {\rm{         }} \to \int {\mathop \prod \limits_I d\pi ^I \exp \left[ {-i12c\pi ^2 \left( x \right)e\left( x \right)d^4 x + i\left( {U_J^I(x  + dx,x )\tilde e^J (x + dx) - \tilde e^I (x)} \right)\pi _I (x)d^3 X} \right]}.  \\
 \end{array}
\]
The exterior covariant derivative
\[
\left(  {U_J^I(x  + dx,x )\tilde e^J (x + dx) - \tilde e^I (x)} \right)d^3 X = D_0\tilde e^I (x) dX^0 d^3 X +O(dx^6) = D\tilde e^I (x)d^3 X+O(dx^6) 
\]
is along the direction of time $dX^0$; the direction of the normal to the surface $\delta M(X^1, X^2, X^3)$. So, it leads to propagation from one surface to another.\\
\\
Thus, we write the amplitude as
\begin{align*}
&\left\langle {\tilde e_I \left( {x + dx} \right)} \right|e^{-ic12\hat \pi ^2 \hat ed^4  x} \left| {\tilde e^I \left( x \right)} \right\rangle _{\delta M_1  \to \delta M_2 }
\\
 &=\int {\mathop \prod \limits_I d\pi ^I \exp \left[ {-i12c\pi ^2 \left( x \right)e\left( x \right)d^4 x + i\pi _I (x)D\tilde e^I (x)d^3 X} \right]}.
\end{align*}
\\
The path integral is the integral of ordered product of those amplitudes on all not intersected 3D closed surfaces, thus we write it as
\begin{align*}
W &= \int {\mathop \prod  \limits_I D\tilde e^I D\pi _I \exp i\int {\left( {-12c\pi ^2 ed^4 x + \pi _I D\tilde e^I d^3 X} \right)} }  \\
&= \int {\mathop \prod  \limits_I D\tilde e^I D\pi _I \exp i\int {\left( {-12c\pi ^2 e^0  \wedge e^1  \wedge e^2  \wedge e^3  + \pi _I D\tilde e^I d^3 X} \right)} }.
\end{align*}
Thus, we obtained the same action \ref{eq:4}. There is no problem with Lorentz non-invariance in $\frac{{\partial \tilde e^I (x)}}{{\partial X^0 }}d^3 XdX^0$, because the equation of motion we get from this path integral is
\[
\frac{{\partial \tilde e^I (x)}}{{\partial X^0 }} \propto  - \pi ^I,
\]
so 
\[
\frac{{\partial \tilde e^I (x)}}{{\partial X^0 }}\pi _I d^3 XdX^0  \propto  - \pi _I \pi ^I d^3 XdX^0.
\]
This is Lorentz invariant. It is similar to equation of motion of scalar field $\phi$; $\pi  = \partial _0 \phi$, which solves the same problem.\\
In our gauge, we have
\begin{align*}
 \pi _I \pi ^I d^3 XdX^0  \to \pi ^2 dX^0  \wedge dX^1  \wedge dX^2  \wedge dX^3  &= \pi ^2 e_\mu ^0 e_\nu ^1 e_\rho ^2 e_\sigma ^3 dx^\mu   \wedge dx^\nu   \wedge dx^\rho   \wedge dx^\sigma   \\
 {\rm{   }} &= \pi ^2 e_\mu ^0 e_\nu ^1 e_\rho ^2 e_\sigma ^3 \varepsilon ^{\mu \nu \rho \sigma } d^4 x = \pi ^2 ed^4 x,
\end{align*}
it is an invariant element.\\

The path integral
\begin{equation}\numberwithin{equation}{section}\label{eq:5}
W = \int {\mathop \prod \limits_I D\tilde e^I D\pi _I \exp i\int {\left( {-12c\pi ^2 e^0  \wedge e^1  \wedge e^2  \wedge e^3  + \pi _I D\tilde e^I d^3 X} \right)} }
\end{equation}
vanishes unless
\begin{align*}
\frac{\delta S(\pi, e)}{{\delta \pi _I }}&=\frac{\delta }{{\delta \pi _I }}\left( {-12c\pi ^2 e^0  \wedge e^1  \wedge e^2  \wedge e^3  + \pi _I D\tilde e^I d^3 X} \right)\\
& = -24c\pi ^I e^0  \wedge e^1  \wedge e^2  \wedge e^3  + D\tilde e^I d^3 X = 0.
\end{align*}
Therefore we get equation of motion of $\pi^I$
\begin{equation}\numberwithin{equation}{section}\label{eq:6}
 \pi ^I  = \frac{{  1}}{{24c}}\left( { e^0  \wedge  e^1  \wedge  e^2  \wedge  e^3 } \right)^{ - 1} {D}{ {\tilde e}} {^I} d^3 X,
\end{equation}
or
\begin{equation}\numberwithin{equation}{section}
\pi ^I \pi ^J  = \frac{1}{{\left( {24c} \right)^2 }}\left( {e^0  \wedge e^1  \wedge e^2  \wedge e^3 } \right)^{ - 2} D\tilde e^I d^3 X D\tilde e^J d^3 X.
\end{equation}
So,
\begin{align*}
 -12c\pi ^2& e^0  \wedge e^1  \wedge e^2  \wedge e^3  + \pi _I D\tilde e^I d^3 X = \frac{-1}{{48c}}\left( {e^0  \wedge e^1  \wedge e^2  \wedge e^3 } \right)^{ - 1} \left( {D\tilde e_I d^3 X} \right)\left( {D\tilde e^I d^3 X} \right) \\
 {\rm{  }} &+ \frac{1}{{24c}}\left( {e^0  \wedge e^1  \wedge e^2  \wedge e^3 } \right)^{ - 1} \left( {D\tilde e_I d^3 X} \right)\left( {D\tilde e^I d^3 X} \right).
 \end{align*}
Inserting it in the path integral \ref{eq:5}, we get
\begin{equation}\numberwithin{equation}{section}\label{eq:7}
W = \int {\mathop \prod \limits_I D\tilde e^I exp} \frac{{  i}}{{48c}}\int {\left( {e^0  \wedge e^1  \wedge e^2  \wedge e^3 } \right)^{ - 1} \left( {D\tilde e_I d^3 X} \right)\left( {D\tilde e^I d^3 X} \right)}.
\end{equation}
Now, we calculate it using the gravitational field $ e^I$. The canonical field $\tilde e^I$ is defined in
\[
\tilde e^K d^3 X = e^{K\mu } e\varepsilon _{\mu \nu \rho \sigma } dx^\nu   \wedge dx^\rho   \wedge dx^\sigma  /3!.
\]
Applying the covariant exterior derivative, we get
\[
\left( {D {\tilde e}{^K} } \right)d^3 X = \left( { D_{\mu _1 }  e^{K\mu } } \right) e\varepsilon _{\mu \nu \rho \sigma } d x^{\mu _1 }  \wedge d x^\nu   \wedge d x^\rho   \wedge d x^\sigma  /3!,
\]
where the covariant derivative D is defined in
\[
DV^I  = dV^I  + \omega {^I} {_J} \wedge V^J.
\]
So the term
\begin{equation}\numberwithin{equation}{section}\label{eq:ba}
\left( {e^0  \wedge e^1  \wedge e^2  \wedge e^3 } \right)^{ - 1} \left( {D\tilde e_I d^3 X} \right)\left( {D\tilde e^I d^3 X} \right) = \frac{{\left( {D\tilde e_I d^3 X} \right)\left( {D\tilde e^I d^3 X} \right)}}{{e^0  \wedge e^1  \wedge e^2  \wedge e^3 }},
\end{equation}
in the path integral \ref{eq:7}, becomes
\begin{equation}\numberwithin{equation}{section}\label{eq:ba1}
 \frac{{\left( { D_{\mu _1 }  e_I^\mu  } \right) e\varepsilon _{\mu \nu \rho \sigma } d x^{\mu _1 }  \wedge d x^\nu   \wedge d x^\rho   \wedge dx^\sigma  \left( { D_{\mu _2 }  e^{I\mu '} } \right) e\varepsilon _{\mu '\nu '\rho '\sigma '} d x^{\mu _2 }  \wedge d x^{\nu '}  \wedge d x^{\rho '}  \wedge d x^{\sigma '} }}{{3!3! e_{\mu _3 }^0  e_{\nu _3 }^1  e_{\rho _3 }^2  e_{\sigma _3 }^3 d x^{\mu _3 }  \wedge d x^{\nu _3 }  \wedge d x^{\rho _3 }  \wedge d x^{\sigma _3 } }}.
\end{equation}
The terms \ref{eq:ba} and \ref{eq:ba1} contain only 4-forms, therefore the result is 4-form, thus its component must be totally anti-symmetric. To get this result, we do the following procedures.\\

Let us write the contraction $\frac{dx^\mu}{dx^\nu }=\delta _{\nu} ^\mu $ as $(dx^\mu, \partial_\nu )=\delta _{\nu} ^\mu $, and $(d^3 x^\mu, \partial^3_\nu )=\delta _{\nu} ^\mu $, with it we can define inversion as
\[
\left( {e_\mu ^0 e_\nu ^1 e_\rho ^2 e_\sigma ^3 dx^\mu   \wedge dx^\nu   \wedge dx^\rho   \wedge dx^\sigma  } \right)^{ - 1}  = E_0^{\mu '} E_1^{\nu '} E_2^{\rho '} E_3^{\sigma '} \frac{\partial }{{\partial x^{\sigma '} }} \wedge \frac{\partial }{{\partial x^{\rho '} }} \wedge \frac{\partial }{{\partial x^{\nu '} }} \wedge \frac{\partial }{{\partial x^{\mu '} }}.
\]
We can rewrite:
\[
e_\mu ^0 e_\nu ^1 e_\rho ^2 e_\sigma ^3 dx^\mu   \wedge dx^\nu   \wedge dx^\rho   \wedge dx^\sigma   = \frac{1}{4}ed^3 x_\mu   \wedge dx^\mu.
\]
(Actually, we have to rewrite the tensors $\varepsilon ^{\mu \nu \rho \sigma } $ and $\varepsilon _{\mu \nu \rho \sigma } $ as $e^{ - 1} \varepsilon ^{\mu \nu \rho \sigma } $ and ${\rm{ }}e\varepsilon _{\mu \nu \rho \sigma } $, but here we neglect this because we get same results).\\
\\
Also, we can rewrite:
\[
E_0^{\mu '} E_1^{\nu '} E_2^{\rho '} E_3^{\sigma '} \partial _{\sigma '}  \wedge \partial _{\rho '}  \wedge \partial _{\nu '}  \wedge \partial _{\mu '}  = E\partial _\nu   \wedge \partial ^{3\nu } ,
\]
with index contracting like
\[
\left( {E\partial _\nu   \wedge \partial ^{3\nu } } \right)\left( {\frac{1}{4}ed^3 x_\mu   \wedge dx^\mu  } \right) = \frac{1}{4}Ee\partial _\nu   \wedge \partial ^{3\nu } d^3 x_\mu   \wedge dx^\mu   = \frac{1}{4}Ee\left( {\delta _\mu ^\nu  } \right)\partial _\nu  dx^\mu   = Ee = 1.
\]
In general, we can write it as
\[
\left( {E\partial _\nu   \wedge \partial ^{3\nu } } \right)\left( {ed^3 x_{\mu'}   \wedge dx^\mu  } \right) = Ee\partial _\nu   \wedge \partial ^{3\nu } d^3 x_{\mu'}   \wedge dx^\mu   =Ee {\delta _{\mu'} ^\nu  } \partial _\nu  dx^\mu   = {\delta _{\mu'} ^\mu  } ,
\]
or
\[
\left( {E   \partial ^{3\nu } \wedge  \partial _\nu } \right)\left( {e dx^\mu  \wedge  d^3 x_{\mu'} } \right) = Ee    \partial ^{3\nu }\wedge \partial _\nu   dx^\mu  \wedge d^3 x_{\mu'}     =Ee {\delta ^{\mu} _\nu  } \partial ^{3\nu }  d^3 x_{\mu'} = {\delta _{\mu'} ^\mu  } ,
\]
In the path integral term eq.(1.15), let us make the replacements:
\[
\left( {D_{\mu _1 } e_I^\mu  } \right)e\varepsilon _{\mu \nu \rho \sigma } dx^{\mu _1 }  \wedge dx^\nu   \wedge dx^\rho   \wedge dx^\sigma  /3! \to \left( {D_{\mu _1 } e_I^\mu  } \right)edx^{\mu _1 }  \wedge d^3 x_\mu   =  - \left( {D_{\mu _1 } e_I^\mu  } \right)ed^3 x_\mu   \wedge dx^{\mu _1 },
\]
and
\[
\left( {D_{\mu _2 } e^{I\mu '} } \right)e\varepsilon _{\mu '\nu '\rho '\sigma '} dx^{\mu _2 }  \wedge dx^{\nu '}  \wedge dx^{\rho '}  \wedge dx^{\sigma '} /3! \to  - \left( {D_{\mu _2 } e^{I\mu '} } \right)ed^3 x_{\mu '}  \wedge dx^{\mu _2 } .
\]
Let us assume the following replacement:
\[
d^3 x_\mu   \wedge dx^\mu   =  - dx_\mu   \wedge d^3 x^\mu   \to d^3 x_\mu   \wedge dx^{\mu _1 }  =  - dx_\mu   \wedge d^3 x^{\mu _1 } .
\]
There is no problem with this trick because in any 4D spacetime we have the contraction $(d^3 x_\mu   \wedge dx^{\nu })=\delta _{\mu }^\nu d^4x$.\\
\\
Therefore, we make the replacement:
\[
 - \left( {D_{\mu _1 } e_I^\mu  } \right)ed^3 x_\mu   \wedge dx^{\mu _1 }  \to \left( {D_{\mu _1 } e_I^\mu  } \right)edx_\mu   \wedge d^3 x^{\mu _1 }.
\]
By that, the term
\[
 \frac{{\left( { D_{\mu _1 }  e_I^\mu  } \right) e\varepsilon _{\mu \nu \rho \sigma } d x^{\mu _1 }  \wedge d x^\nu   \wedge d x^\rho   \wedge dx^\sigma  \left( { D_{\mu _2 }  e^{I\mu '} } \right) e\varepsilon _{\mu '\nu '\rho '\sigma '} d x^{\mu _2 }  \wedge d x^{\nu '}  \wedge d x^{\rho '}  \wedge d x^{\sigma '} }}{{3!3! e_{\mu _3 }^0  e_{\nu _3 }^1  e_{\rho _3 }^2  e_{\sigma _3 }^3 d x^{\mu _3 }  \wedge d x^{\nu _3 }  \wedge d x^{\rho _3 }  \wedge d x^{\sigma _3 } }},
\]
in the path integral \ref{eq:7}, becomes
\[
 - \left( {E\partial _\nu   \wedge \partial ^{3\nu } } \right)\left( {\left( {D_{\mu _1 } e_I^\mu  } \right)edx_\mu   \wedge d^3 x^{\mu _1 } } \right)\left( {\left( {D_{\mu _2 } e^{I\mu '} } \right)ed^3 x_{\mu '}  \wedge dx^{\mu _2 } } \right)\\
\]
\[
=\left( {D^{\mu _1 } e_{I\mu } } \right)\left( {D_{\mu _2 } e^{I\mu '} } \right)e\left( {\partial _\nu   \wedge \partial ^{3\nu } } \right)\left( {d^3 x_{\mu _1 }  \wedge dx^\mu  } \right)\left( {d^3 x_{\mu '}  \wedge dx^{\mu _2 } } \right),
\]
where we used
\[
-dx_\mu   \wedge d^3 x^{\mu _1 }=d^3 x^{\mu _1 }\wedge dx_\mu  \text{ }then\text{ } d^3 x_{\mu _1 }\wedge dx^\mu.
\]
Thus we can write
\[
\frac{{\left( {D\tilde e_I d^3 X} \right)\left( {D\tilde e^I d^3 X} \right)}}{{e^0  \wedge e^1  \wedge e^2  \wedge e^3 }} \to \left( {D^{\mu _1 } e_{I\mu } } \right)\left( {D_{\mu _2 } e^{I\mu '} } \right)e\left( {\partial _\nu   \wedge \partial ^{3\nu } } \right)\left( {d^3 x_{\mu _1 }  \wedge dx^\mu  } \right)\left( {d^3 x_{\mu '}  \wedge dx^{\mu _2 } } \right).
\]
Let us choose the contraction:
\[
\left( {\partial _\nu   \wedge \partial ^{3\nu } } \right)\left( {d^3 x_{\mu _1 }  \wedge dx^\mu  } \right)\left( {d^3 x_{\mu '}  \wedge dx^{\mu _2 } } \right) = \left( {\partial _\nu   \wedge \partial ^{3\nu } d^3 x_{\mu _1 }  \wedge dx^\mu  } \right)\left( {d^3 x_{\mu '}  \wedge dx^{\mu _2 } } \right)
\]
\[
 = \delta _{\mu _1 }^\nu  \left( {\partial _\nu   \wedge dx^\mu  } \right)\left( {d^3 x_{\mu '}  \wedge dx^{\mu _2 } } \right) = \delta _{\mu _1 }^\nu  \left( { - dx^\mu   \wedge \partial _\nu  } \right)\left( { - dx^{\mu _2 }  \wedge d^3 x_{\mu '} } \right)
\]
\[
 = \delta _{\mu _1 }^\nu  dx^\mu   \wedge \partial _\nu  dx^{\mu _2 }  \wedge d^3 x_{\mu '}  = \delta _{\mu _1 }^\nu  \delta _\nu ^{\mu _2 } dx^\mu   \wedge d^3 x_{\mu '} .
\]
So, in the path integral \ref{eq:7}, we make the replacement:
\[
\frac{{\left( {D\tilde e_I d^3 X} \right)\left( {D\tilde e^I d^3 X} \right)}}{{e^0  \wedge e^1  \wedge e^2  \wedge e^3 }} \to \left( {D^{\mu _1 } e_{I\mu } } \right)\left( {D_{\mu _2 } e^{I\mu '} } \right)e\delta _{\mu _1 }^\nu  \delta _\nu ^{\mu _2 } dx^\mu   \wedge d^3 x_{\mu '}
\]
\[
 = \left( {D_\nu  e_{I\mu } } \right)\left( {D^\nu  e^{I\mu '} } \right)edx^\mu   \wedge d^3 x_{\mu '}  =  - \left( {D_\nu  e_{I\mu } } \right)\left( {D^\nu  e^{I\mu '} } \right)ed^3 x_{\mu '}  \wedge dx^\mu
\]
\[
 =  - \left( {D_\nu  e_{I\mu } } \right)\left( {D^\nu  e^{I\mu '} } \right)e\delta _{\mu '}^\mu  d^4 x =  - \left( {D_\nu  e_{I\mu } } \right)\left( {D^\nu  e^{I\mu } } \right)ed^4 x.
\]
We can also choose another contraction:
\[
\left( {D^{\mu _1 } e_{I\mu } } \right)\left( {D_{\mu _2 } e^{I\mu '} } \right)e\left( {\partial _\nu   \wedge \partial ^{3\nu } } \right)\left( {d^3 x_{\mu _1 }  \wedge dx^\mu  } \right)\left( {d^3 x_{\mu '}  \wedge dx^{\mu _2 } } \right) \to
\]
\[
\text{\text{\text{\text{ } } } }  \left( {D^{\mu _1 } e_{I\mu } } \right)\left( {D_{\mu _2 } e^{I\mu '} } \right)e\left( {\partial _\nu   \wedge \partial ^{3\nu } d^3 x_{\mu _1 }  \wedge dx^\mu  } \right)\left( {d^3 x_{\mu '}  \wedge dx^{\mu _2 } } \right)
\]
\[
 \text{\text{\text{\text{ } } } } = \left( {D^{\mu _1 } e_{I\mu } } \right)\left( {D_{\mu _2 } e^{I\mu '} } \right)e\left( {\delta _{\mu _1 }^\nu  \partial _\nu  dx^\mu  } \right)\left( {d^3 x_{\mu '}  \wedge dx^{\mu _2 } } \right)
\]
\[
\text{\text{\text{\text{ } } } } = \delta _{\mu _1 }^\nu  \delta _\nu ^\mu  \left( {D^{\mu _1 } e_{I\mu } } \right)\left( {D_{\mu _2 } e^{I\mu '} } \right)e\left( {d^3 x_{\mu '}  \wedge dx^{\mu _2 } } \right).
\]
So we get
\[
\frac{{\left( {D\tilde e_I d^3 X} \right)\left( {D\tilde e^I d^3 X} \right)}}{{e^0  \wedge e^1  \wedge e^2  \wedge e^3 }} \to \left( {D^\mu  e_{I\mu } } \right)\left( {D_{\mu '} e^{I\mu '} } \right)ed^4 x.
\]
\\
Considering the two possible contractions, we write final result as
\[
-\left( {e^0  \wedge e^1  \wedge e^2  \wedge e^3 } \right)^{ - 1} \left( {D\tilde e_I d^3 X} \right)\left( {D\tilde e^I d^3 X} \right) = \frac{{ 1}}{2}\left( {D_\mu  e_I^\nu  D^\mu  e_\nu ^I  - D_\mu  e_I^\nu  D_\nu  e^{I\mu } } \right)ed^4 x.
\]
This Lagrangian can be written as 4-form as required. It depends only on covariant derivative of the gravitational field $e^I$, similarly to Lagrangian of electromagnetic field in curved spacetime. It is also independent of the gauge we chose for the surface $\delta M$.\\
This Lagrangian is invariant under local Lorentz transformation $V^I  \to L{^I}{_J} (x)V^J $ and under any coordinate transformation $V^\mu   \to \frac{{\partial x^\mu  }}{{\partial x'^\nu  }}V'^\nu$.\\

The path integral of the gravitational field
\[
W = \int {\mathop \prod \limits_I D\tilde e^I exp} \frac{{  i}}{{48c}}\int {\left( {e^0  \wedge e^1  \wedge e^2  \wedge e^3 } \right)^{ - 1} \left( {D\tilde e_I d^3 X} \right)\left( {D\tilde e^I d^3 X} \right)}
\]
becomes
\[
W  = \int {\mathop \prod \limits_I De^I \exp \frac{i}{{48c}}\frac{1}{2}\left( { - D_\mu  e_I^\nu  D^\mu  e_\nu ^I  + D_\mu  e_I^\nu  D_\nu  e^{I\mu } } \right)ed^4 x} ,
\]\\
with gravitational field Lagrangian:
\begin{equation}\numberwithin{equation}{section}\label{eq:b9}
Ld^4 x = \frac{1}{{48c}}\frac{1}{2}\left( { - D_\mu  e_I^\nu  D^\mu  e_\nu ^I  + D_\mu  e_I^\nu  D_\nu  e^{I\mu } } \right)ed^4 x.
\end{equation}\\
The covariant derivative $D$ is defined as $DV^I=dV^I+\omega{^I}{_J}\wedge V^J$. This path integral is now defined on the states $\left| { e^{I}_\mu } \right\rangle$ and $\left| {\omega^{IJ}_\mu } \right\rangle$, but with considering that it is integrated over $\omega^{IJ}_\mu$. So, the equation of motion of $\omega^{IJ}_\mu$ must be satisfied in this Lagrangian. We determine the constant $c$ using the Newtonian gravitational potential $c \succ 0$.
\\

Comparing \ref{eq:6} and \ref{eq:7} with \ref{eq:b9}, we get 
\begin{equation}\numberwithin{equation}{section}\label{eq:b6}
\pi^2=\frac{1}{{(24c)^2}}\frac{1}{2}\left( { - D_\mu  e_I^\nu  D^\mu  e_\nu ^I  + D_\mu  e_I^\nu  D_\nu  e^{I\mu } } \right),
\end{equation} 
which determines $\pi^2$ for given $e^I$ and $\omega^{IJ}$.
\\

In weak gravity, we can use the background spacetime approximation: ${D_\mu   \to \partial _\mu  }$ and $e \to 1 + \delta e$, so we get
\[
L \to \frac{1}{{48c}}\frac{1}{2}\left( { - \partial _\mu  e_I^\nu  \partial ^\mu  e_\nu ^I  + \partial _\mu  e_I^\nu  \partial _\nu  e^{I\mu } } \right)
\]
or
\[
L_0  = \frac{1}{{48c}}\frac{1}{2}\eta _{IJ} e_\mu ^I \left( {g^{\mu \nu } \partial ^2  - \partial ^\mu  \partial ^\nu  } \right)e_\nu ^J .
\]
But, in strong gravity, we can not use this approximation. So, we have a problem with the determinant $e$ in the path integral
\[
W= \int {\mathop \prod \limits_I De^I exp} \frac{i}{{48c}}\int {\frac{1}{2}\left( { - D_\mu  e_I^\nu  D^\mu  e_\nu ^I  + D_\mu  e_I^\nu  D_\nu  e^{I\mu } } \right)e_{\mu _1 }^0 e_{\nu _1 }^1 e_\rho ^2 e_\sigma ^3 \varepsilon ^{\mu _1 \nu _1 \rho \sigma } d^4 x}.
\]
Or
\[
 \int {\mathop \prod \limits_I De^I exp} \frac{i}{{48c}}\int {\frac{1}{2}\left( { - D_\mu  e_I^\nu  D^\mu  e_\nu ^I  + D_\mu  e_I^\nu  D_\nu  e^{I\mu } } \right)ed^4 x} .
\]
The path integral is independent of arbitrary changing of the coordinates $x^{\mu}$, this changes the determinant $e \to e'= e+\delta e$, then we have 
\[
\frac{\delta S}{\delta e}=0,
\]
 it yields
\[
- D_\mu  e_I^\nu  D^\mu  e_\nu ^I  + D_\mu  e_I^\nu  D_\nu  e^{I\mu } =0 \to D_\mu  e^{I\nu}=0.
\]
So,
\[
 \text{ }\text{ }L \left(\omega, e\right)= 0,\text{ }\text{ then}\text{ }\text{ }H \left(\omega, e\right)= 0.
\]
This path integral is trivial; there is no propagation because there is no gravitational energy: $H \left(\omega, e\right)=0$. The reason is that the gravitational field $e_\mu ^I $ has the entity of spacetime. It is impossible for spacetime to be dynamical on itself, to propagate over itself.\\

But, if we use the approximation $e_\mu ^I (x) \to \delta _\mu ^I  + h_\mu ^I (x)$, the path integral exists. So, the propagation is possible. Thus, the gravitational field propagates freely only on background spacetime. This is case of weak gravity at low energy densities.\\

In background spacetime, we set ${g = \eta }$ and $k_\mu  e^{\mu I}  = 0$, so the path integral of weak gravitational field becomes
\begin{equation}\numberwithin{equation}{section}
W = \int {\mathop \prod \limits_I De^I \exp } i\int {\frac{1}{{48c}}\frac{1}{2}e_\mu ^I \left( {\eta _{IJ} g^{\mu \nu } \partial ^2  - \eta _{IJ} \partial ^\mu  \partial ^\nu  } \right)e_\nu ^J d^4 x} .
\end{equation}
Thus, the free gravitational field propagator becomes
\[
\Delta _{IJ }^{\mu \nu } (x_2  - x_1 ) = 48c\int {\frac{{d^4 k}}{{(2\pi )^4 }}\frac{{\eta_{IJ } g^{\mu \nu } e^{ik(x_2  - x_1 )} }}{{k^2  - i\varepsilon }}},
\]
or
\begin{equation}\numberwithin{equation}{section}
\Delta _{\rho \sigma }^{\mu \nu } (x_2  - x_1 ) = 48c\int {\frac{{d^4 k}}{{(2\pi )^4 }}\frac{{g_{\rho \sigma } g^{\mu \nu } e^{ik(x_2  - x_1 )} }}{{k^2  - i\varepsilon }}} .
\end{equation}
We will use this propagation in gravitational field interaction with scalar and spinor fields.\\

Let us write the Lagrangian \ref{eq:b9} using Riemann curvature. Integrate it by parts, we get
\[
Ld^4 x \to \frac{1}{{48c}}\frac{1}{2}\left( {   e_I^\nu  D^\mu D_\mu e_\nu ^I  -  e{_I}^\nu D_\mu D_\nu  e^{I\mu } } \right)ed^4 x,
\]
and using $D_\mu D_\nu=[D_\mu, D_\nu]+D_\nu D_\mu$, we get
\[
Ld^4 x \to \frac{1}{{48c}}\frac{1}{2}\left( {   e_I^\nu  D^\mu D_\mu e_\nu ^I  -  e{_I}^\nu [D_\mu, D_\nu] e^{I\mu }-e{_I}^\nu D_\nu D_\mu  e^{I\mu } } \right)ed^4 x.
\]
But, $[D_\mu, D_\nu] e^{I\mu }={(R_{\mu\nu})^I}_{J} e^{J\mu }$, where ${(R_{\mu\nu})^I}_{J}$ is Riemann curvature tensor. So
\[
Ld^4 x \to \frac{1}{{48c}}\frac{1}{2}\left( {   e_I^\mu  D^2 e_\mu ^I  -e{_I}^\mu D_\mu D_\nu  e^{I\nu} +  e^{I\mu} e^{J \nu}{(R_{\mu\nu})}_{IJ}   } \right)ed^4 x.
\]
The term $e^{I\mu} e^{J \nu}{(R_{\mu\nu})}_{IJ}e $ is usual Lagrangian of general relativity which gives Einstein equation in vacuum. 
\begin{equation}\numberwithin{equation}{section}\label{eq:b11}
\frac{\delta }{\delta e_I^\mu}\int {e^{I\mu} e^{J \nu}{(R_{\mu\nu})}_{IJ}ed^4x}\to R^I_\mu-\frac{1 }{2}Re^I_\mu=0.
\end{equation}
In order to get same equation, we set
\[
\frac{\delta }{\delta e_I^\mu}( e_I^\mu   D^\rho D_\rho e_\mu ^I  -e{_I}^\mu D_\mu D_\nu  e^{I\nu})= D^\rho  D_\rho e_\mu ^I  - D_\mu D_\nu  e^{I\nu}=0.
\]
It is satisfied by choosing ${D_\mu}e^{I} _\nu=0$ which relates the connection with $e^{I} _\nu$. 
\\

Now, we simplify the formula \ref{eq:3}:
\begin{equation}\numberwithin{equation}{section}\label{eq:b12}
 c\varepsilon _{IJKL} e^I  \wedge e^J  \wedge\left(d\Omega +dB+\Omega\wedge B+B\wedge\Omega+B\wedge B\right)^{KL}=\pi _I D\tilde e^I d^3 X
\end{equation}
by using the equation \ref{eq:6} of the momentum $\pi^I$:
\[
 \pi ^I  = \frac{{  1}}{{24c}}\left( { e^0  \wedge  e^1  \wedge e^2  \wedge  e^3 } \right)^{ - 1} {D}{ {\tilde e}} {^I} d^3 X.
\]
Omitting ${D}{ {\tilde e}} {^I} d^3 X$ from both equations, we get
\begin{align*}
 \varepsilon _{IJKL} e^I  \wedge e^J  \wedge\left(d\Omega +dB+\Omega\wedge B+B\wedge\Omega+B\wedge B\right)^{KL}&=24\pi^2\left(e^0  \wedge e^1  \wedge e^2  \wedge e^2  \right)\\
&=24\pi^2 ed^4x.
\end{align*}
From \ref{eq:1a} and \ref{eq:2a}, we get
\[
 \varepsilon _{IJKL} e^I  \wedge e^J  \wedge\left(\Omega\wedge\Omega \right)^{KL}=-12\pi^2 ed^4x.
\]
Using it in the previous formula, we get
\begin{align*}
 \varepsilon _{IJKL} e^I  \wedge e^J  \wedge (d\Omega +dB+\Omega\wedge B&+B\wedge\Omega+B\wedge B)^{KL}\\
&=-2\varepsilon _{IJKL} e^I  \wedge e^J  \wedge\left(\Omega\wedge\Omega \right)^{KL}.
\end{align*}
Or
\begin{equation}\numberwithin{equation}{section}
 \varepsilon _{IJKL} e^I  \wedge e^J  \wedge (d\Omega +dB+\Omega\wedge B+B\wedge\Omega+B\wedge B+2\Omega\wedge\Omega)^{KL}=0,
\end{equation}
this formula determines the tensor $\Omega$ as a function of $e^I$ and $B^{IJ}$. \\

We need to show that there is at least one solution for this equation. Let us choose
\begin{equation}\numberwithin{equation}{section}\label{eq:b8}
 \varepsilon _{IJKL} e^I  \wedge e^J  \wedge (dB+B\wedge B)^{KL}=-Fed^4 x,
\end{equation}
where $F$ is a scalar function. Therefore eq.(1.20) becomes
\[
 \varepsilon _{IJKL} e^I  \wedge e^J  \wedge (d\Omega +\Omega\wedge B+B\wedge\Omega+2\Omega\wedge\Omega )^{KL}=Fed^4 x,
\]
multiplying with 2, it becomes
\[
 \varepsilon _{IJKL} e^I  \wedge e^J  \wedge (d2\Omega +2\Omega\wedge B+B\wedge2\Omega+2\Omega\wedge2\Omega )^{KL}=2Fed^4 x,
\]
and adding eq.(1.21) again, we obtain
\[
 \varepsilon _{IJKL} e^I  \wedge e^J  \wedge (d(2\Omega+B) +(2\Omega+B)\wedge (2\Omega+B) )^{KL}=Fed^4 x.
\]
For $2\Omega+B=\omega'$, we have
\begin{equation}\numberwithin{equation}{section}\label{eq:b7}
 \varepsilon _{IJKL} e^I  \wedge e^J  \wedge (d\omega' +\omega'\wedge \omega' )^{KL}=Fed^4 x.
\end{equation}
The solutions of \ref{eq:b8} and \ref{eq:b7} are same that of the formula \ref{eq:3} with \ref{eq:6}.\\

For given $e^I $ and $B^{IJ}$, we obtain $F$ from \ref{eq:b8} and by using it in \ref{eq:b7} we obtain $\omega'^{IJ}$. Therefore we obtain the tensor $\Omega^{IJ}=\left(\omega'-B\right)^{IJ}/2$ from which we calculate the conjugate momentum $\pi ^I$ by using ${\Omega }^{IJ}= \pi ^{IJK} e_K$ and $\pi ^{IJK}  = \pi _L \varepsilon ^{LIJK}$. And so obtaining the connection \ref{eq:c4}, $\omega^{IJ}=\Omega^{IJ}+B^{IJ}$. Using \ref{eq:6}, we obtain $D\tilde e^I d^3 X$.  \\
 
For given $e^I $ and the connection $\omega^{IJ}$ in the Lagrangian \ref{eq:b9}, we determine the tensor $\Omega^{IJ}$ by using the conjugate momentum $\pi ^I$ which can be determined from $\pi^2$, where we use the given $e^I $ and $\omega^{IJ}$ in \ref{eq:b6}:
\[
\pi^2=\frac{1}{{(24c)^2}}\frac{1}{2}\left( { - D_\mu  e_I^\nu  D^\mu  e_\nu ^I  + D_\mu  e_I^\nu  D_\nu  e^{I\mu } } \right).
\]
We use $\omega=\Omega+B$ in $\omega'=2\Omega+B$ to get $\omega'=\Omega+\omega$, then use $\omega'$ in \ref{eq:b7} to get $F$ which can be used in \ref{eq:b8} to determine the connection $B$. Therefore from known $e^I $ and $\omega^{IJ}$, we get the conjugate momentum $\pi ^I$ and the connection $B$ which satisfy \ref{eq:3} with \ref{eq:6}. Therefore there is at least one solution for the formula \ref{eq:3} with $e^I $ and $\omega^{IJ}$ are given.

\section{ Lagrangian of the Plebanski two form field}
 The Plebanski two form complex field ${\Sigma^i}$, in selfdual representation $\left| {\Sigma^i} \right\rangle $, is defined by $\Sigma ^{i}  =P^{i}_{IJ} \Sigma^{IJ}$, where $\Sigma^{IJ}=e^I \wedge e^J$ is real anti-symmetric two form and $ {P^i }$ is selfdual projector given in[7, 8]
\[
\text{ \text{\text{\text{ } } } }\left( {P^i } \right)_{jk}  = \frac{1}{2}\varepsilon {^i} _{jk} {\rm{  \text{ }\text{, }  }}\left( {P^i } \right)_{0j}  = \frac{i}{2}\delta ^i _j {\rm{ \text{ }\text{: } }}i = I{\rm{ }}\text{ }\text{for }{\rm{ }}I = 1,2,3.
\]
That is
\[
\Sigma^i=\frac{1}{2}\varepsilon {^i} _{jk} e^j \wedge e^k +ie^0 \wedge e^i
\]
The complex field ${\Sigma^i}$ has spatial Lorentz index $i=I=1, 2, 3$, so it transforms under $SO(3)$, the subgroup of Lorentz group $SO(3,1)$. We derive its Lagrangian with dependence only on the covariant derivative of it, then we search for conditions which satisfy reality of the Lagrangian. We start with Lagrangian of the gravitational field eq.(1.6):
\[
 S(e, \omega )= c\int \varepsilon _{IJKL} e^I  \wedge e^J  \wedge \left( d\Omega +dB+\Omega\wedge \Omega+\Omega\wedge B+B\wedge\Omega+B\wedge B\right)^{KL}.
\]
As we did before, we try to find Lagrangian that transforms covariantly under Local Lorentz transformations and contains only ${\Sigma^i}$, and cancel out the remaining terms. For that, we separate the action $S(e, \omega )$ to $S(e, \Omega )+S(e, \Omega\wedge B, B )$. Because the actions $S(e, \omega )$ and $S(e, \Omega )$ transform covariantly, the action $S(e, \Omega\wedge B, B )$ also transforms covariantly. Therefore, we can choose $S(e, \Omega\wedge B, B )=0$.\\

We get Lagrangian with terms $ ({^*}D \Sigma _{i}) \wedge D\Sigma ^{i}$ and $ ({^*}D \Sigma _{IJ}) \wedge D\Sigma ^{IJ}$ similarly to electromagnetic field. With a metric $g_ {\mu\nu}$ on $M$.\\

Let us write this action as
\[
\int \varepsilon _{IJKL} e^I  \wedge e^J  \wedge \left( d\Omega +\Omega\wedge \Omega\right)^{KL}+I(e,  \Omega\wedge B, B ).
\]
Using the assumption $\Omega^{IJ}=\pi {_M}{ ^{IJ}} e^M$ in the first part, we obtain
\begin{align*}
&S(e,\pi ) =
\\
& c\int {\left[ {\varepsilon _{IJKL} e^I  \wedge e^J  \wedge d\left( {\pi {_M}{ ^{KL}} e^M } \right) + \varepsilon _{IJKL} e^I  \wedge e^J  \wedge \left( {\pi {_{K_1 }}{^K}{ _M} } \right)e^{K_1 }  \wedge \left( {\pi {_{K_2 }}{ ^{ML}} } \right)e^{K_2 } } \right]}\\
&+I(e,  \Omega\wedge B, B ).
\end{align*}
We assume that the integral of
\[
\varepsilon _{IJKL} d\left( {e^I  \wedge e^J  \wedge \left( {\pi {_M} ^{KL} e^M } \right)} \right) = \varepsilon _{IJKL} d\left( {\Sigma ^{IJ}  \wedge \left( {\pi {_M }^{KL} e^M } \right)} \right)
\]
is zero at infinities. Using
\[
d\Sigma ^{IJ}  \wedge \left( {\pi {_M} ^{KL} } \right)e^M  + e^I  \wedge e^J  \wedge d\left( {\pi {_M} ^{KL} e^M } \right) =  - \left( {\pi {_M} ^{KL} } \right)e^M  \wedge d\Sigma ^{IJ}  + e^I  \wedge e^J  \wedge d\left( {\pi {_M} ^{KL} e^M } \right),
\]
 the action becomes
\begin{align*}
S(e,\pi ) =& c\int {\left[ {\varepsilon _{IJKL} \left( {\pi {_M} ^{KL} } \right)e^M  \wedge d\Sigma ^{IJ}  + \varepsilon _{IJKL} \Sigma ^{IJ}  \wedge \left( {\pi {_{K_1 }}{^K }{ _M}} \right)\left( {\pi {_{K_2 }} ^{ML} } \right)e^{K_1 }  \wedge e^{K_2 } } \right]}\\
 &+I(e, \Omega\wedge B, B ),
\end{align*}
or
\begin{align*}
S(e,\pi ) =& c\int {\left[ {\varepsilon _{IJKL} \left( {\pi {_M }^{KL} } \right)e^M  \wedge d\Sigma ^{IJ}  + \varepsilon _{IJKL} \left( {\pi {_{K_1 }} {^K }{_M}} \right)\left( {\pi {_{K_2} } ^{ML} } \right)\Sigma ^{IJ}  \wedge \Sigma ^{K_1 K_2 } } \right]}\\
&+I(e, \Omega\wedge B, B ).
\end{align*}
Using our assumption:
\[
\pi ^{IJK}  = \pi _L \varepsilon ^{LIJK} ,
\]
we get
\[
\text{ }\text{ }\text{ }\text{ }\varepsilon _{IJKL} \left( {\pi {_M }^{KL} } \right)e^M  = \varepsilon _{IJKL} \pi ^{MKL} e_M  = \varepsilon _{IJKL} \pi _N \varepsilon ^{NMKL} e_M  =  - 2\left( {\pi _I e_J  - \pi _J e_I } \right).
\]
Making the replacement:
\[
\Sigma ^{IJ}  \wedge \Sigma ^{K_1 K_2 }  \to \varepsilon ^{IJK_1 K_2 } \Sigma ^{01}  \wedge \Sigma ^{23},
\]
 we get
\begin{align*}
\varepsilon _{IJKL} \left( {\pi {_{K_1 }}{^K} {_M} } \right)&\left( {\pi {_{K_2 }} ^{ML} } \right)\Sigma ^{IJ}  \wedge \Sigma ^{K_1 K_2 }  = \varepsilon _{IJKL} \left( {\pi {_{K_1 }}{^K} {_M} } \right)\left( {\pi {_{K_2 }} ^{ML} } \right)\varepsilon ^{IJK_1 K_2 } \Sigma ^{01}  \wedge \Sigma ^{23} \\
&= 2\left( {\pi {_{L }}{^K} {_M} } \right)\left( {\pi {_K} ^{ML} } \right)\Sigma ^{01}  \wedge \Sigma ^{23}  = 2\left( {\pi _{LKM} } \right)\left( {\pi ^{KML} } \right)\Sigma ^{01}  \wedge \Sigma ^{23} \\
& = 2\left( {\pi _{KML} } \right)\left( {\pi ^{KML} } \right)\Sigma ^{01}  \wedge \Sigma ^{23}  = 2\pi ^I \varepsilon _{IKML} \pi _J \varepsilon ^{JKML} \Sigma ^{01}  \wedge \Sigma ^{23} \\
& =  - 12\pi ^2 \Sigma ^{01}  \wedge \Sigma ^{23} .
\end{align*}
Therefore, the action becomes
\begin{equation}\label{eq:25}
S(e,\pi ,\Sigma ) = c\int {\left[ { - 2\left( {\pi _I e_J  - \pi _J e_I } \right) \wedge d\Sigma ^{IJ}  - 12\pi _I \pi ^I \Sigma ^{01}  \wedge \Sigma ^{23} } \right]}+I(e, \Omega\wedge B, B ) .
\end{equation}
Because the real Plebanski two form $\Sigma ^{IJ}=e^I \wedge e^J$ is anti-symmetric, and $\Sigma ^{01}  \wedge \Sigma ^{23}=ed^4x$, we can rewrite:
\[
S(e,\pi ,\Sigma ) = c\int {\left[ { - 4\pi _I e_J  \wedge d\Sigma ^{IJ}  - 12\pi _I \pi ^I ed^4x } \right]}+I(e, \Omega\wedge B, B ).
\]
Now we let $\Pi _{IJ}=(\pi _I e_J -\pi _J e_I)/2$ which can be conjugate to $\Sigma ^{IJ}$ in $3+1$ decomposition. We write
\begin{equation}\label{eq:21}
S(e,\pi ,\Sigma ) = c\int {\left[ { - 4\Pi _{IJ}  \wedge d\Sigma ^{IJ}  - 12\pi _I \pi ^I ed^4x } \right]}+I(e, \Omega\wedge B, B ).
\end{equation}
Using self-dual projection properties
\begin{equation}\label{eq:20}
 P_i^{IJ}P^i_{KL} +{\bar P}_i^{IJ}{\bar P}^i_{KL}=\frac{1}{2}(\delta^I_K \delta^J_L-\delta^I_L \delta^J_K), \text{ and } { P}^i_{IJ}{\bar P}^{IJ}_k=0,
\end{equation}
we obtain 
\[
\Pi _{IJ}  \wedge d\Sigma ^{IJ}=\Pi _{i}  \wedge d\Sigma ^{i}+C.C, 
\]
with 
\[
\Pi _{i}= P_i^{IJ}\Pi _{IJ} \text{ and } \Sigma ^{i}=P^i_{IJ}\Sigma ^{IJ}.
\]
Using it in Lagrangian \ref{eq:21}, it becomes
\begin{equation}\label{eq:22}
S(e,\pi ,\Sigma ) = c\int {\left( { - 4\Pi _{i}  \wedge d\Sigma ^{i}+C.C  - 12\pi _I \pi ^I ed^4x } \right)}+I(e, \Omega\wedge B, B ).
\end{equation}
We need to write $\pi _I \pi ^I$ using these variables, we have
\[
\Pi _{i\mu}= P_i^{IJ}\Pi _{IJ\mu}= P_i^{IJ}(\pi _I e_{J\mu} -\pi _J e_{I\mu})/2=P_i^{IJ}\pi _I e_{J\mu},
\]
therefore by using self-dual projection properties \ref{eq:20}, we obtain
\[
\Pi _{i\mu}\Pi ^{i\mu}+C.C=P_i^{IJ}\pi _I e_{J\mu}P^i_{I'J'}\pi ^{I'} e^{J'\mu}+C.C=P_i^{IJ}P^i_{I'J'}\pi _I e_{J\mu}\pi ^{I'} e^{J'\mu}+C.C,
\]
but $\pi ^I $ and $e_{\mu}^I$ are real, so
\[
\Pi _{i\mu}\Pi ^{i\mu}+C.C=(P_i^{IJ}P^i_{I'J'}+C.C)\pi _I \pi ^{I'}e_{J\mu} e^{J'\mu}=\frac{1}{2}(\delta^I_{I'} \delta^J_{J'}-\delta^I_{J'} \delta^J_{I'})\pi _I \pi ^{I'}e_{J\mu} e^{J'\mu},
\]
then using $e_{J\mu} e^{J'\mu}=\delta_J^{J'}$ and $e_{J\mu} e^{J\mu}=4$, it becomes
\[
\Pi _{i\mu}\Pi ^{i\mu}+C.C=2\pi _I \pi ^{I}-    \frac{1}{2}\pi _I \pi ^{J}e_{J\mu} e^{I\mu}=2\pi _I \pi ^{I}-    \frac{1}{2}\pi _I \pi ^{I}=\frac{3}{2}\pi ^{2}.
\]
Therefore we write the Lagrangian \ref{eq:22} as
\begin{equation}\label{eq:23}
S(\Pi ,\Sigma ) = c\int {\left( { - 4\Pi _{i}  \wedge d\Sigma ^{i} -8\Pi _{i\mu}\Pi ^{i\mu} ed^4x+C.C  } \right)}+I(e, \Omega\wedge B, B ).
\end{equation}
If we add the term $\Delta S$, which by it we get
\begin{equation}\label{eq:24}
S(\Pi ,\Sigma ) = c\int {\left( { - 4\Pi _{i}  \wedge D\Sigma ^{i} -8\Pi _{i\mu}\Pi ^{i\mu} ed^4x+C.C  } \right)}+\Delta S+I(e, \Omega\wedge B, B ).
\end{equation}
With covariant derivative regarding the connection $A^i=P^i_{IJ} \omega^{IJ}$.
\\

In this formula we regard the scalar term $\Pi _{i\mu}\Pi ^{i\mu}$ as energy density of $\Sigma ^{i}$, while the form $\Pi _{i}  \wedge d\Sigma ^{i} $ as kinetic energy of it. We can choose $\Delta S+I(e, \Omega\wedge B, B )=0$. Thus we obtain self-dual action
\[
S_{self-dual}(\Pi ,\Sigma ) = -8c\int {\left( {  \frac{1}{2}\Pi _{i}  \wedge D\Sigma ^{i} +\Pi _{i\mu}\Pi ^{i\mu} ed^4x } \right)}.
\]
On $3+1$ decomposition of the manifold $M$, we get the term $\frac{1}{2}\varepsilon ^{a0bc}\Pi _{ia}  D_0\Sigma ^{i}_{bc}ed^4x=-\frac{1}{2}\varepsilon ^{0abc}\Pi _{ia}  D_0\Sigma ^{i}_{bc}ed^4x$ in which we define the gravitational electric field $E^{ia}=\varepsilon ^{0abc}\Sigma ^{i}_{bc}$ with conjugate momentum $\Pi _{ia}$.
\\

We can remove $\Pi ^{i}$ from this Lagrangian, $\frac{\delta}{\delta \Pi}S_{self-dual}=0$, and keep only $D\Sigma ^{i}$ which contians only $\Sigma ^{i}$ and the connection $\omega$ in $D=d+\omega$, as we did for $e ^{I}$. Thus
\[
L_{self-dual}(\Pi ,\Sigma ) =\frac{8c}{16} {    D_\mu \Sigma _{i\nu\rho} \left(D^\mu \Sigma ^{i\nu\rho}- D^\nu \Sigma ^{i\mu\rho}+...   \right)ed^4x } ,
\]
or
\[
L_{self-dual}(\Pi ,\Sigma ) =\frac{8c}{16}     ({^*}D \Sigma _{i}) \wedge D\Sigma ^{i}  ,
\]
where ${^*}$ is hudge operator with respect to a metric $g_{\mu\nu}$ on $M$.
\\

We can treat the action \ref{eq:25} in different way. Using $\varepsilon _{0123}  =  - 1$, we rewrite it as
\[
S(e,\pi ,\Sigma ) = c\int {\left[ { - 4\pi _I e_J  \wedge d\Sigma ^{IJ}  + 12\pi _I \pi ^I \varepsilon _{IJKL} \Sigma ^{IJ}  \wedge \Sigma ^{KL} /4!} \right]}+ I(e, \Omega\wedge B, B) ,
\]
or
\[
S(e,\pi ,\Sigma ) = c\int {\left[ { - 4\pi _I e_J  \wedge d\Sigma ^{IJ}  + \frac{1}{2}\pi ^2 \varepsilon _{IJKL} \Sigma ^{IJ}  \wedge \Sigma ^{KL} } \right]}+ I(e, \Omega\wedge B, B ).
\]
The path integral over the momentum $\pi^I$ vanishes unless 
\[
\frac{\delta S(e, \pi, \Sigma)}{\delta \pi_I}=\frac{\delta }{{\delta \pi _I }}\int {\left[ { - 4\pi _I e_J  \wedge d\Sigma ^{IJ}  + \frac{1}{2}\pi ^2 \varepsilon _{IJKL} \Sigma ^{IJ}  \wedge \Sigma ^{KL} } \right]} + \frac{\delta I(e,  \Omega\wedge B, B)}{\delta \pi_I}= 0.
\]
So we get the equation of motion of $\Sigma ^{IJ}$.
But, it is not easy to separate $\Sigma$ from $e$. It is similar to the gravitational field, it is separable only in weak gravity in background spacetime. Thus, we solve it in background spacetime, then we generate the solution to arbitrary spacetime. Therefore, we use the approximation:
\[
\int {\left( { - 4\pi _I e_J  \wedge d\Sigma ^{IJ}  + \frac{1}{2}\pi ^2 \varepsilon _{IJKL} \Sigma ^{IJ}  \wedge \Sigma ^{KL} } \right)}
\]
\[
\to\int {\left( { - 4\pi _I e_{\mu J} \partial _\nu  \Sigma _{\rho \sigma }^{IJ} \varepsilon ^{\mu \nu \rho \sigma }  + \frac{1}{2}\pi ^2 \varepsilon _{IJKL} \Sigma _{\mu \nu }^{IJ} \Sigma _{\rho \sigma }^{KL} \varepsilon ^{\mu \nu \rho \sigma } } \right)} d^4 x.
\]
 The background spacetime approximation is
\[
\text{ \text{\text{\text{ } } } }e_\mu ^I (x) \to \delta _\mu ^I  + h_\mu ^I (x){\rm{ }}\text{ \text{\text{\text{, } } } }{\rm{ }}e \to 1 + \delta e,
\]
thus we get
\[
\Sigma _{\mu \nu }^{IJ}  = \frac{1}{2}\left( {e_\mu ^I e_\nu ^J  - e_\nu ^I e_\mu ^J } \right)
\to \frac{1}{2}\left( {\delta _\mu ^I \delta _\nu ^J  - \delta _\nu ^I \delta _\mu ^J } \right) + \frac{1}{2}\left( {h_\mu ^I \delta _\nu ^J  - h_\nu ^I \delta _\mu ^J } \right) + \frac{1}{2}\left( {\delta _\mu ^I h_\nu ^J  - \delta _\nu ^I h_\mu ^J } \right).
\]
Inserting it in the action
\[
{\rm{     }}S\left( {e,\Sigma } \right) =c \int {\left( { - 4\pi _I e_{\mu J} \partial _\nu  \Sigma _{\rho \sigma }^{IJ} \varepsilon ^{\mu \nu \rho \sigma }  + \frac{1}{2}\pi ^2 \varepsilon _{IJKL} \Sigma _{\mu \nu }^{IJ} \Sigma _{\rho \sigma }^{KL} \varepsilon ^{\mu \nu \rho \sigma } } \right)} d^4 x+I(e, \Omega\wedge B, B ),
\]
it becomes
\[
S\left( {e,\Sigma } \right) \to S\left( {h,\delta \Sigma } \right) =c \int {\left( { - 4\pi _I \partial _\nu  \Sigma _{\rho \sigma }^{IJ} \varepsilon {_J} ^{\nu \rho \sigma }  + \frac{1}{2}\pi ^2 \left( { - 24} \right) +  \ldots } \right)} d^4 x .
\]
Therefore, the condition
\[
\frac{\delta }{{\delta \pi _I }}\int {\left[ { - 4\pi _I e_J  \wedge d\Sigma ^{IJ}  + \frac{1}{2}\pi ^2 \varepsilon _{IJKL} \Sigma ^{IJ}  \wedge \Sigma ^{KL} } \right]}  + \frac{\delta I(e,  \Omega\wedge B, B)}{\delta \pi_I}= 0
\]
approximates to
\[
\frac{\delta }{{\delta \pi _I }}\int {\left( { - 4\pi _I \partial _\nu  \Sigma{^I} _{J\rho \sigma } \varepsilon ^{J\nu \rho \sigma }  + \frac{1}{2}\pi ^2 \left( { - 24} \right)} +...\right)} d^4 x = 0.
\]
Its solution is
\[
\pi ^I  =  - \frac{1}{6}\partial _\nu  \Sigma _{J\rho \sigma }^I \varepsilon ^{J\nu \rho \sigma } +... =  - \frac{1}{6}\partial ^\nu  \Sigma ^{IJ\rho \sigma } \varepsilon _{J\nu \rho \sigma } +...
\]
Thus, the action in the background spacetime approximates to
\[
S(\Sigma ) \to c\int {\left[ {\frac{2}{3}\partial ^{\nu _1 } \Sigma ^{IJ_1 \rho _1 \sigma _1 } \varepsilon _{J_1 \nu _1 \rho _1 \sigma _1 } \partial _\nu  \Sigma _{IJ\rho \sigma } \varepsilon ^{J\nu \rho \sigma }  + ...} \right]d^4 x} .
\]
Defining inner product via $\Sigma ^{IJ_1 \rho _1 \sigma _1 } \Sigma _{IJ\rho \sigma }  = \Sigma ^2 \delta _J^{J_1 } \delta _\rho ^{\rho _1 } \delta _\sigma ^{\sigma _1 } $, we get
\[
S(\Sigma ) \to c\int {\left( { - 4\partial _\mu  \Sigma _{IJ}^{\nu \rho } \partial ^\mu  \Sigma _{\nu \rho }^{IJ}  + ...} \right)d^4 x} {\rm{    }}\text{ \text{\text{\text{ with} } } }{\rm{   }}\partial _\mu  \Sigma _{IJ}^{\mu \rho }  = 0.
\]
This is action of real Plebanski two form in approximation of background spacetime. It is similar to scalar field. The corresponding Lagrangian is
\[
L_0 (\Sigma ) \to  - 4c\left( {\partial _\mu  \Sigma _{IJ}^{\nu \rho } } \right)\left( {\partial ^\mu  \Sigma _{\nu \rho }^{IJ} } \right){\rm{    }}\text{\text{\text{\text{ with} } } }{\rm{   }}\partial _\mu  \Sigma _{IJ}^{\mu \rho }  = 0.
\]
In curved spacetime, we rewrite it as
\begin{equation}\numberwithin{equation}{section}
L_0 (\Sigma )d^4 x \rightarrow -4c'\left( {\partial _\mu  \Sigma _{IJ}^{\nu \rho } } \right)\left( {\partial^\mu  \Sigma _{\nu \rho }^{IJ} } \right)ed^4 x .
\end{equation}
It does not transform covariantly because the partial derivative $\partial_\mu$ does not. But the total Lagrangian $L(e, \omega )$ transforms covariantly, thus
$L(e, \omega )= L(e, d\Sigma, \Sigma )+L(e, \Sigma\wedge B, B )$ transforms covariantly, Let us rewrite:
 \[
L(e, d\Sigma,\Sigma )+L(e, \Sigma\wedge B, B )=L(e, d\Sigma, \Sigma )+\Delta L -\Delta L+L(e, \Sigma\wedge B, B ), 
\]
with $L(e, d\Sigma, \Sigma )+\Delta L$ transforms covariantly, and choose $-\Delta L+L(e, \Sigma\wedge B, B )=0$, which determines the reference connection $B$. Thus, we get
\[
L(e, d\Sigma, \Sigma )+\Delta L \to L (\Sigma )d^4 x=-4c'\left( {D _\mu  \Sigma _{IJ}^{\nu \rho } } \right)\left( {D^\mu  \Sigma _{\nu \rho }^{IJ} } \right)ed^4 x,
\]
which transforms covariantly, where $D\Sigma ^{IJ}=d\Sigma ^{IJ}+\omega{^I}{_K}\wedge\Sigma ^{KJ}+\omega{^J}{_K}\wedge\Sigma ^{IK}$.

We get Lagrangian of the complex Plebanski two form field $\Sigma ^{i}  =P^{i}_{IJ} \Sigma^{IJ}$ by using the selfdual projector $P^{i}$, which projects the real Plebanski two form $\Sigma^{IJ}$ into two states: selfdual $\left| {\Sigma^i} \right\rangle $ and anti-selfdual $\left| {\bar \Sigma ^{i} } \right\rangle$.
Thus, the term 
\[
D _\mu  \Sigma _{IJ}^{\nu \rho }  {D^\mu  \Sigma _{\nu \rho }^{IJ} }
\]
 in the Lagrangian becomes 
\[
{D _\mu  \Sigma _{i}^{\nu \rho } } {D^\mu  \Sigma _{\nu \rho }^{i} }+{D _\mu  \bar\Sigma _{i}^{\nu \rho } } {D^\mu  \bar\Sigma _{\nu \rho }^{i} },
\]
where the hermitian conjugate ${{\bar{\Sigma}} _{i}^{\nu \rho } {\bar{\Sigma}} _{\nu \rho }^{i} }$ is represented in anti-selfdual ${\bar{\Sigma}} ^{i}={\bar{P}}_{IJ}^{i}{\Sigma}^{IJ}$.\\
\\
We search for conditions which allow us to rewrite the complex Plebanski two form field ${\Sigma} ^{i}$ as a real field. To do this, let us choose $\bar\Sigma _{i}=0$, which cancells out the terms of $\bar\Sigma _{i}$ and makes $\Sigma _{i}$ real field. So,
\begin{equation}
 {\bar\Sigma}^{i}=\frac{1}{2}\varepsilon ^{ijk} \Sigma_{jk}  - i\Sigma ^{0i}  = 0 \to \frac{1}{2}\varepsilon ^{ijk} \Sigma _{jk}  = i\Sigma ^{0i},
\end{equation}
generally it becomes $\frac{1}{2}\varepsilon ^{IJKL} \Sigma _{KL}  = i\Sigma ^{IJ}$. Therefore, $\Sigma ^{0i}$ and $e^{0}$ are pure imaginary, so we replace $X^0$ by $iX^0$ and the metric $\eta ^{IJ}=(- + + +)$ by $(+ + + +)$.

Therefore, the Plebanski two form field in selfdual representation becomes
\[
 {\Sigma}^{i}=\frac{1}{2}\varepsilon ^{ijk} \Sigma _{jk}  + i\Sigma ^{0i}=\varepsilon ^{ijk} \Sigma _{jk},
\]
which is real as required for satisfying the reality condition. \\
Therefore, the Lagrangian of Plebanski two form field in selfdual representation becomes:
\begin{equation}
L_0 (\Sigma )d^4 x = -4c'\left( {D _\mu  \Sigma _{i}^{\nu \rho } } \right)\left( {D^\mu  \Sigma _{\nu \rho }^{i} } \right)ed^4 x .
\end{equation}
Because $\Sigma _{i}$ is real, so the covairant derivative is $D\Sigma ^{i}=d\Sigma ^{i}+\omega{^i}{_j}\wedge\Sigma ^{j}$.
\\
Let us combine the gravitational and Plebanski fields in one field $K_{\mu} ^i$ via
\[
K_{\mu} ^i  = \frac{1}{2}\left( {e_{\mu} ^i  + \frac{i}{4}\varepsilon ^{ijk} \varepsilon _{0\mu \rho \sigma } \Sigma _{jk}^{\rho \sigma } } \right),
\]
its hermitian conjugate is
\[
\bar K_\mu ^i  = \frac{1}{2}\left( {e_{\mu} ^i  - \frac{i}{4}\varepsilon ^{ijk} \varepsilon _{0\mu \rho \sigma } \Sigma _{jk}^{\rho \sigma } } \right),
\]
 where $i, j$ and $k$ are local-Lorentz frame indices for $I=i=1, 2, 3$. And $K_{\mu} ^0=e_{\mu} ^0$ with $\bar K_{\mu} ^0=-e_{\mu} ^0$, here $e_{\mu} ^0$ is pure imaginary as mentioned in eq.(2.2). Therefore, we get
\[
 K_{\mu }^i  + \bar K_{\mu} ^i  = e_{\mu} ^i  \text{ } \text{and } \text{ }K_{\mu} ^i  - \bar K_{\mu }^i  = \frac{i}{4}\varepsilon ^{ijk} \varepsilon _{0\mu \rho \sigma } \Sigma _{jk}^{\rho \sigma }= \frac{i}{2} \varepsilon _{0\mu \rho \sigma } \Sigma^{i\rho \sigma },    
\]
then we get
\[
\left( {D ^\nu  K_{\mu} ^i  + D^\nu  \bar K_{\mu} ^i } \right)\left( {D _\nu  K_{i}^\mu   + D _\nu  \bar K_{i}^\mu  } \right) = D ^\nu  e_{\mu} ^i D _\nu  e_{i}^\mu , 
\]
and
\begin{align*}
 \left( {D ^\nu  K_{\mu} ^i  - D ^\nu  \bar K_{\mu} ^i } \right)\left( {D_\nu  K_{i}^\mu   - D _\nu  \bar K_{i}^\mu  } \right) &= \frac{{ - 1}}{4}\varepsilon _{0\mu \rho \sigma } \varepsilon _{0} ^{\mu \rho '\sigma '} D ^\nu  \Sigma ^{i\rho \sigma } D _\nu  \Sigma _{i\rho '\sigma '}  \\ 
  &= \frac{1}{4}\varepsilon _{0\mu \rho \sigma } \varepsilon ^{0\mu \rho '\sigma '} D ^\nu  \Sigma ^{i\rho \sigma } D _\nu  \Sigma _{i\rho '\sigma '} \\
& = -D ^\nu  \Sigma ^{i\rho \sigma } D _\nu  \Sigma _{i\rho \sigma }.
\end{align*}
Therefore, we have
\[
D _\nu  e_{i}^\mu D ^\nu  e_{\mu} ^i  +  D_\nu \Sigma ^{i\rho \sigma }  D^\nu \Sigma _{i\rho \sigma }\to 4D _\mu  \bar K^{i} D ^\mu  K_{i}.
\]
Using this in gravitational Lagrangian: 
\[
L\left({e}\right)d^4 x = -\frac{{1}}{{48c}}\frac{1}{2}\left( {D _\mu  e_{I}^\nu  } \right)\left( {D ^\mu  e_\nu ^I } \right)ed^4 x,
\]
and in Plebanski Lagrangian eq.(2.3):  
\[
L \left( {\bar \Sigma } \right)d^4 x = -8c'\frac{1}{2}\left( {D ^\nu  \Sigma ^{i\rho \sigma } D_\nu  \Sigma _{i\rho \sigma } } \right)ed^4 x,
\]
 setting $8c'=1/(48c)$, we get
\begin{equation}\numberwithin{equation}{section}
L\left({e}\right)d^4 x+L \left( {\bar \Sigma } \right)d^4 x \to {\frac{1}{12c}}\frac{-1}{2}\left( {D ^\nu  \bar K_{\mu} ^i D _\nu  K_{i}^\mu  } +{D ^\nu  \bar K_{\mu} ^0 D _\nu  K^{0\mu}  }\right) ed^4 x.
\end{equation}
using the metric $(+ + + +)$, it becomes
 \begin{equation}\numberwithin{equation}{section}
L\left({e}\right)d^4 x+L \left( {\bar \Sigma } \right)d^4 x \to {\frac{1}{12c}}\frac{-1}{2}\left( \delta_{IJ}{D ^\nu  \bar K_{\mu} ^I D _\nu  K^{J\mu}  } \right) ed^4 x.
\end{equation}
Generally, we write
\[
L\left({K}\right)d^4 x= {\frac{1}{12c}}\frac{-1}{2}\delta_{IJ}\left( {D^\nu  \bar K_{\mu} ^I D_\nu  K^{J\mu}  }-{D^\nu  \bar K_{\mu} ^I D^\mu  K^{J}_{\nu}  } \right)ed^4 x.
\]
It satisfies the reality condition as required.\\

\section{Static potential of weak gravity}
We derive the static potential of scalar and spinor fields interactions with weak gravitational field in the static limit; the Newtonian gravitational potential. We find that this potential has same structure for both fields, it depends only on the distances between the particles and on their energies. By that, we determine the constant $c\succ0$.\\

The action of scalar field in arbitrary curved spacetime is[1]
\[
S(e,\phi ) = \int {d^4 xe\left( {\eta ^{IJ} e_I^\mu  e_J^\nu  D_\mu  \phi ^ +  D_\nu  \phi  - V(\phi )} \right)} {\rm{    }}.
\]
In weak gravity, we use approximation of background spacetime:
\[
e_I^\mu  (x) \to \delta _I^\mu   + h_I^\mu  (x)\text{ }\text{ ,}\text{ }\text{ }e \to 1 + \delta e.
\]
Thus, action approximates to
\[
S(e,\phi ) = \int {d^4 x\left( {\partial _\mu  \phi ^ +  \partial ^\mu  \phi  + h^{\mu \nu } (x)\partial _\mu  \phi ^ +  \partial _\nu  \phi  + h^{\nu \mu } (x)\partial _\mu  \phi ^ +  \partial _\nu  \phi  - V(\phi ) + ...} \right)} {\rm{    }}.
\]
The gravitational field is symmetric, so we get
\[
S(e,\phi ) = \int {d^4 x\left( {\partial _\mu  \phi ^ +  \partial ^\mu  \phi  + 2h^{\mu \nu } (x)\partial _\mu  \phi ^ +  \partial _\nu  \phi  - V(\phi ) + ...} \right)} {\rm{    }}.
\]
The energy-momentum tensor of scalar field is[9]
\[
T_{\mu \nu }  = \partial _\mu  \phi ^ +  \partial _\nu  \phi  + g_{\mu \nu } L{\rm{   }},
\]
hence
\[
\partial _\mu  \phi ^ +  \partial _\nu  \phi  = T_{\mu \nu }  - g_{\mu \nu } L{\rm{   }}.
\]
Inserting it in the Lagrangian, it becomes
\[
L = \partial _\mu  \phi ^ +  \partial ^\mu  \phi  + 2h^{\mu \nu } (x)\left( {T_{\mu \nu }  - g_{\mu \nu } L} \right) - V(\phi ) + ...{\rm{   }}.
\]
or
\[
L = \partial _\mu  \phi ^ +  \partial ^\mu  \phi  + 2h^{\mu \nu } T_{\mu \nu }  - V(\phi ) - 2h^{\mu \nu } g_{\mu \nu } L + ...{\rm{  }}.
\]
Therefore, in the interaction term, we make the replacement:
\[
\partial _\mu  \phi ^ +  \partial _\nu  \phi  \to T_{\mu \nu } {\rm{   }}\text{ }\text{  }and\text{ }\text{ }{\rm{   }}V \to V + 2h^{\mu \nu } g_{\mu \nu } L.
\]
Because the gravitational field is weak, $2h^{\mu \nu } g_{\mu \nu } L$ is neglected compared with $L$.\\

We find the potential $V(r)$ of exchanged virtual gravitons by two particles $k_1$ and $k_2$ using $M\left( {k_1  + k_2  \to k'_1  + k'_2 } \right)$ matrix element (like Born approximation to the scattering amplitude in non-relativistic quantum mechanics [10]).\\
\\
For one of Feynman diagrams, we have
\[
 \text{ }\text{ }\text{ }\text{ }\text{ }\text{ }\text{ }\text{ }iM\left( {k_1  + k_2  \to k'_1  + k'_2 } \right) = i\left( { - ik'_2 } \right)_\mu  \left( {ik_2 } \right)_\nu  \frac{{\bar \Delta ^{\mu \nu \rho \sigma } \left( q \right)}}{i}i\left( { - ik'_1 } \right)_\rho  \left( {ik_1 } \right)_\sigma ,
\]
with
\[
\text{ }\text{ }\text{ }\text{ }{\rm{   }}q = k'_1  - k_1  = k_2  - k'_2 .
\]
The propagator $\Delta ^{\mu \nu \rho \sigma } \left( {x_2  - x_1 } \right)$ is the gravitons propagator eq.(1.18) that we get from Lagrangian of free gravitational field in background spacetime:
\[
L_0  = \frac{1}{{48c}}\frac{1}{2}\eta _{IJ} e_\mu ^I \left( {g^{\mu \nu } \partial ^2  - \partial ^\mu  \partial ^\nu  } \right)e_\nu ^J  \to \frac{1}{{48c}}\frac{1}{2}\eta _{IJ} h_\mu ^I \left( {g^{\mu \nu } \partial ^2  - \partial ^\mu  \partial ^\nu  } \right)h_\nu ^J .
\]
With the gauge $\partial ^\mu  e_\mu ^I  = 0$, we get
\[
\Delta _{\mu \nu }^{IJ} \left( {y - x} \right) = \int {\frac{{d^4 q}}{{\left( {2\pi } \right)^4 }}} \bar \Delta _{\mu \nu }^{IJ} \left( {q^2 } \right)e^{iq(y - x)} \text{ }\text{ }\text{: }\text{ }\bar \Delta _{\mu \nu }^{IJ} \left( {q^2 } \right) = 48c\frac{{g_{\mu \nu } \eta ^{IJ} }}{{q^2  - i\varepsilon }}.
\]
Therefore, the $M$ matrix element becomes
\[
 \text{ }\text{ }\text{ }\text{ }\text{ }\text{ }\text{ }\text{ }iM\left( {k_1  + k_2  \to k'_1  + k'_2 } \right) = i48c\left( { - ik'_2 } \right)_\mu  \left( {ik_2 } \right)_\rho  \frac{{g^{\mu \nu } g^{\rho \sigma } }}{{q^2 }}\left( { - ik'_1 } \right)_\sigma  \left( {ik_1 } \right)_\nu   ,
\]
where
\[
 g = \eta {\rm{   }}\text{ }\text{and }{\rm{   }}q = k'_1  - k_1  = k_2  - k'_2 . 
\]
Comparing it with[10]
\[
iM\left( {k_1  + k_2  \to k'_1  + k'_2 } \right) =  - i\bar V\left( q \right)\delta ^4 \left( {k_{out}  - k_{in} } \right),
\]
we get
\[
\bar V\left( {q^2 } \right) =  - 48c\left( { - ik'_2 } \right)_\mu  \left( {ik_2 } \right)_\rho  \frac{{g^{\mu \nu } g^{\rho \sigma } }}{{q^2 }}\left( { - ik'_1 } \right)_\sigma  \left( {ik_1 } \right)_\nu  .
\]
Then, comparing this formula with the replacement:
\[
\partial _\mu  \phi ^ +  \partial _\nu  \phi  \to T_{\mu \nu },
\]
and evaluating inverse Fourier transform, we get
\[
V\left( {y - x} \right) =  - 48cT_{\mu \rho } \left( y \right)g^{\mu \nu } g^{\rho \sigma } T_{\nu \sigma } \left( x \right)\frac{1}{{4\pi \left| {y - x} \right|}} =  - 48c\frac{{T_{\mu \nu } \left( y \right)T^{\mu \nu } \left( x \right)}}{{4\pi \left| {y - x} \right|}},
\]
where $T^{\mu \nu }$ is transferred energy-momentum tensor. It is anti-symmetric, so the summation over the indices ${\mu}$ and ${\nu }$ is repeated twice. Therefore, we divide the right side by 2:
\[
V\left( {y - x} \right) = - {\frac{48c}{2}}\frac{{T_{\mu \nu } \left( y \right)T^{\mu \nu } \left( x \right)}}{{4\pi \left| {y - x} \right|}}.
\]

In static limit, for one particle, we approximate $T^{00}$ to $m$, where $m$ is the mass of interacted particles.\\
Thus, we get Newtonian gravitational potential:
\[
V\left( {y - x} \right) =  - {\frac{48c}{2}}\frac{{m^2 }}{{4\pi \left| {y - x} \right|}} =  - G\frac{{m^2 }}{{\left| {y - x} \right|}} \to 48c = 8\pi G.
\]
Therefore, the weak gravitational Lagrangian becomes
\[
L_0  = \frac{1}{{4\pi G}}{\frac{1}{4}}\eta _{IJ} e_\mu ^I \left( {g^{\mu \nu } \partial ^2  - \partial ^\mu  \partial ^\nu  } \right)e_\nu ^J .
\]

We do the same thing for spinor fields' interactions with gravitational field. The action is[1]
\[
S(e,\psi ) = \int {d^4x e\left( {ie_I^\mu  \bar \psi \gamma ^I D_\mu  \psi  - m\bar \psi \psi } \right)} {\rm{    }},
\]
where the covariant derivative $D_\mu$ is
\[
D_\mu   = \partial _\mu  {\rm{ + }}\left( {\omega _\mu  } \right)_J^I L_I^J {\rm{ + }}A_\mu ^a {\rm{T}}^a {\rm{   }}.
\]
In background spacetime, it becomes
\[
S(e,\psi ) = \int {d^4 x\left( {i\bar \psi \gamma ^\mu  D_\mu  \psi  + ih_I^\mu  \bar \psi \gamma ^I D_\mu  \psi  - m\bar \psi \psi }+ ... \right) } {\rm{    }}.
\]
Let us consider only the terms:
\[
\int {d^4 x\left( {i\bar \psi \gamma ^\mu  \partial _\mu  \psi  + ih_\nu ^\mu  \bar \psi \gamma ^\nu  \partial _\mu  \psi  - m\bar \psi \psi } \right)} {\rm{      }}\text{ }\text{: }\text{ }\text{ }g = \eta .
\]
The energy-momentum tensor of spinor field is[9]
\[
T^{\mu \nu }  =  - i\bar \psi \gamma ^\mu  \partial ^\nu  \psi  + g^{\mu \nu } L.
\]
Therefore, in the interaction term, we have the replacements
\[
i\bar \psi \gamma ^\mu  \partial ^\nu  \psi  \to  - T_{\mu \nu } {\rm{   }}\text{ }\text{  }and\text{ }\text{ }{\rm{   }}L \to L + h^{\mu \nu } g_{\mu \nu } L.
\]
The term $h^{\mu \nu } g_{\mu \nu } L$ is neglected compared with the Lagrangian $L$. We find $M$ matrix element of exchanged virtual gravitons $p_1+p_2 \to p'_1+p'_2$. For one of Feynman diagrams[10]:
\[
iM\left( {p_1  + p_2  \to p'_1  + p'_2 } \right) = i48c\bar u\left( {p'_1 } \right)\gamma ^\mu  \left( { - ip_1 } \right)_\nu  u\left( {p_1 } \right)\frac{{g_{\mu \sigma } g^{\nu \rho } }}{{q^2 }}\bar u\left( {p'_2 } \right)\gamma ^\sigma  \left( { - ip_2 } \right)_\rho  u\left( {p_2 } \right),
\]
with
\[
q = p'_1  - p_1  = p_2  - p'_2 \text{ }\text{and }\text{ }\text{ }g=\eta ,
\]
we get
\[
\bar V\left( {q^2 } \right) =  - 48c\bar u\left( {p'_1 } \right)\gamma ^\mu  \left( { - ip_1 } \right)_\nu  u\left( {p_1 } \right)\frac{{g_{\mu \sigma } g^{\nu \rho } }}{{q^2 }}\bar u\left( {p'_2 } \right)\gamma ^\sigma  \left( { - ip_2 } \right)_\rho  u\left( {p_2 } \right).
\]
Comparing this formula with the replacement:
\[
i\bar \psi \gamma ^\mu  \partial ^\nu  \psi  \to  - T_{\mu \nu },
\]
and evaluating inverse Fourier transform, we get
\[
V\left( {y - x} \right) =  - 48c\left( { - T_{\mu \rho } \left( y \right)} \right)g^{\mu \nu } g^{\rho \sigma } \left( { - T_{\nu \sigma } \left( x \right)} \right)\frac{1}{{4\pi \left| {y - x} \right|}} =  - 48c\frac{{T_{\mu \nu } \left( y \right)T^{\mu \nu } \left( x \right)}}{{4\pi \left| {y - x} \right|}},
\]
where $T^{\mu \nu }$ is transferred energy-momentum tensor. Dividing the right side by 2:
 \[
V\left( {y - x} \right) =  - {\frac{48c}{2}}\frac{{T_{\mu \nu } \left( y \right)T^{\mu \nu } \left( x \right)}}{{4\pi \left| {y - x} \right|}}.
\]

In the static limit, for one particle, we approximate $T^{00}$ to $m$, where $m$ is the mass of interacted particles.\\
Thus, we get Newtonian gravitational potential:
\[
V\left( {y - x} \right) =  - {\frac{48c}{2}}\frac{{m^2 }}{{4\pi \left| {y - x} \right|}} =  - G\frac{{m^2 }}{{\left| {y - x} \right|}} \to 48c = 8\pi G.
\]
It is the same potential we found for the interaction of a scalar field with the gravitational field.

\section{Summary}
We have derived Lagrangian of gravitational field with dependence only on the second covariant derivative, like electromagnetic and scalar fields, that means it has the same symmetries. So it makes it easier for unification the gravity with other fields. We postulated the gravity propagation as expansion of closed 3D surfaces in 4D arbitrary spacetime manifold, so this propagation relates to changing in geometry of those surfaces due to that expansion. This is dynamics of gravity; changing the geometry. We suggested spin connection splitting, $\omega \to \Omega+B$, and canonical states $\left| {\tilde e^I } \right\rangle$ and $\left| {\pi ^I } \right\rangle$ just for using them in path integral to get eq.(1.11):
\[
W = \int {\mathop \prod \limits_I D\tilde e^I D\pi _I \exp i\int {\left( {-12c\pi ^2 e^0  \wedge e^1  \wedge e^2  \wedge e^3  + \pi _I D\tilde e^I d^3 X} \right)} }.
\]
Comparing the Lagrangian ${-12c\pi ^2 e^0  \wedge e^1  \wedge e^2  \wedge e^3  + \pi _I D\tilde e^I d^3 X}$ with Lagrangian eq.(1.6): 
\[
c \varepsilon _{IJKL} e^I  \wedge e^J  \wedge \left( d\Omega +dB+\Omega\wedge \Omega+\Omega\wedge B+B\wedge\Omega+B\wedge B\right)^{KL},
\]
we postulate the equality eq.(1.7):
\[
 c\varepsilon _{IJKL} e^I  \wedge e^J  \wedge\left(d\Omega +dB+\Omega\wedge B+B\wedge\Omega+B\wedge B\right)^{KL}=\pi _I D\tilde e^I d^3 X.
\]
We wrote this formula in simpler form; eq.(1.19).\\
 In that path integral we considered $c\varepsilon _{IJKL} e^I  \wedge e^J  \wedge \left(\Omega\wedge \Omega\right)^{KL}$ as self-energy of $\tilde e^I$ on the surface $\delta M$, and $\pi _I D\tilde e^I d^3 X$ as kinetic energy which relates to expansion of those surfaces. Using the original states  $\left| { e^{I}_\mu } \right\rangle$ and $\left| {\omega^{IJ}_\mu } \right\rangle$, we get
\[
16\pi GL (e,\omega)= \left( { - D_\mu  e_I^\nu  D^\mu  e_\nu ^I  + D_\mu  e_I^\nu  D_\nu  e^{I\mu } } \right)ed^4 x.
\] 
We use same methods for Plebanski field to get Lagrangian like 
\[
L (\Sigma )d^4 x = -4c'\left( {D _\mu  \Sigma _{i}^{\nu \rho } } \right)\left( {D^\mu  \Sigma _{\nu \rho }^{i} } \right)ed^4 x,
\]
 with the gauge $(1/2)\varepsilon ^{IJKL} \Sigma _{KL}  = i\Sigma ^{IJ}$. Therefore, $\Sigma ^{0i}$ and $e^{0}$ are pure imaginary, so we replace $X^0$ by $iX^0$ and the metric $\eta ^{IJ}=(- + + +)$ by $(+ + + +)$. This metric allows us to combine the gravititional and Plebanski fields in one field $K_{\mu} ^I$, we get the Lagrangian 
\[24cL\left({K}\right)d^4 x= -\delta_{IJ}\left( {D^\nu  \bar K_{\mu} ^I D_\nu  K^{J\mu}  }-{D^\nu  \bar K_{\mu} ^I D^\mu  K^{J}_{\nu}  } \right)ed^4 x.\]
 Finally we derived the interaction potential of spinor and scalar fields with the gravity in static limit ''Newtonian gravitational potential'' which allows us to determine the constant $c$.

\end{document}